\documentclass{article}
\usepackage{amsmath,amssymb,color,url,amsthm}
\usepackage[pdftex,hiresbb]{graphicx}
\usepackage{type1cm}
\usepackage{subcaption}
\usepackage{geometry}
\usepackage{here}
\allowdisplaybreaks
\newtheorem{thm}{Theorem}[section]
\newtheorem{cor}[thm]{Corollary}

\newtheorem{prop}[thm]{Proposition}

\newtheorem{rem}[thm]{Remark}
\newtheorem{ass}[thm]{Assumption}
\newtheorem{ex}[thm]{Example}
\def\l     {\left}
\def\r     {\right}
\def\<     {\langle}
\def\>     {\rangle}
\def\fin   {\hfill{$\Box$}\vspace{5mm}}
\def\bbE   {{\mathbb E}}
\def\bbP   {{\mathbb P}}
\def\bbR   {{\mathbb R}}

\def\calF  {{\mathcal F}}
\def\calL  {{\mathcal L}}

\def\olcalL{{\overline \calL}}
\def\vt    {\vartheta}

\def\tN    {\widetilde{N}}
\def\tV    {\widetilde{V}}
\def\whS   {\widehat{S}}
\def\old   {\overline{d}}
\def\olz   {\overline{z}}
\def\olZ   {\overline{Z}}

\begin{document}
\title{Approximate option pricing formula for Barndorff-Nielsen and Shephard model}
\author{Takuji Arai\footnote{Department of Economics, Keio University, 2-15-45 Mita, Minato-ku, Tokyo, 108-8345, Japan. (arai@econ.keio.ac.jp)}}
\maketitle

\begin{abstract}
For the Barndorff-Nielsen and Shephard model, we present approximate expressions of call option prices based on the decomposition formula developed by Arai \cite{A}.
Besides, some numerical experiments are also implemented to make sure how effective our approximations are.
\end{abstract}

%
%
\setcounter{equation}{0}
\section{Introduction}
The Barndorff-Nielsen and Shephard (BNS) model is a representative jump-type stochastic volatility model,
which has been actively researched since it was undertaken by Barndorff-Nielsen and Shephard \cite{BNS1}, \cite{BNS2}, but it is by no means tractable mathematically.
Indeed, there are no closed-form expressions of option prices for the BNS model, although option prices themselves are numerically computable by using e.g. the fast Fourier transform.
Then we aim to derive approximations of call option prices based on the decomposition formula developed by Arai \cite{A}.
Approximations obtained in this paper will be useful to analyze the structure of implied volatilities and to develop calibration procedures for model parameters.

As a similar expression to the so-called Hull-White formula, Al\`os \cite{Alos12} has derived a new decomposition formula of option prices for the Heston model
by making use of Ito calculus.
As its extension to the BNS model, \cite{A} presented a decomposition expression, in which option prices for the BNS model are decomposed into 7 terms:
a primal term, an additional term, and 5 residual terms.
The primal term is given by substituting the present volatility value for the Black-Scholes formula,
and the additional term is given by integration with respect to the L\'evy measure of the jump component in the BNS model,
of which integrand is the sum of a difference and a partial derivative of the Black-Scholes formula.
Since it is very complicated to develop a computational method for the additional term, we shall suggest in this paper two types of correction terms.
One is available for around at the money (ATM) and in the money (ITM) options, and converges to the additional term as the time to maturity tends to 0
with order $1$ for around ATM options and with order $\frac{3}{2}$ for ITM options that are not around ATM.
On the other hand, the other is available for ITM options alone, and its rate of convergence to the additional term is of order $1$.
Moreover, we shall show that all the 5 residual terms converge to 0 as the time to maturity tends to 0 with order $\frac{3}{2}$.
Hence, adding a correction term to the primal term, we obtain an approximate option pricing formula for the BNS model.
Thus, we shall produce three approximation formulas including one hybrid version of the two correction terms.
Remark that we do not assume the finite activity of the jump component in this paper.
Besides, we shall implement some numerical experiments to examine the effectiveness of our approximations.
The results of our experiments indicate that our approximations are effective particularly when the time to maturity is short.

For various stochastic volatility models, some authors have studied approximate option pricing expressions based on Al\`os type decomposition,
which has been undertaken by \cite{Alos12} as mentioned above for the Heston model: a representative continuous stochastic volatility model.
Further, she also in \cite{Alos12} developed two types of approximate option pricing formulas with order $\frac{3}{2}$.
A similar approximation with order $1$ has been presented by Al\`os et al. \cite{ADV15}.
As an extension of \cite{Alos12}, Merino and Vives \cite{MV17} gave a decomposition and an approximation for spot-dependent volatility models including the CEV model.
As for jump type models, Merino et al. \cite{MPSV18} treated jump type stochastic volatility models with finite activity and obtained an approximation formula with order $1$.
To our best knowledge, there is no result on this topic for jump models with infinite activity.

This paper is organized as follows: Some mathematical preliminaries and notations are given in the following section.
Approximation formulas and numerical results are introduced in Section 3.
Section 4 is devoted to proofs of our main results, and Section 5 concludes this paper.

%
%
\setcounter{equation}{0}
\section{Preliminaries}
\subsection{Model description}
Throughout this paper, we consider a financial market model being composed of one risky asset and one riskless asset with interest rate $r\geq0$ and finite time horizon $T>0$.
In the BNS model, the risky asset price at time $t\in[0,T]$ is described by
\[
S_t:=S_0\exp\l\{\int_0^t\l(r+\mu-\frac{1}{2}\Sigma_u^2\r)du+\int_0^t\Sigma_udW_u+\rho H_{\lambda t}\r\}, \ \ \ t\in[0,T],
\]
where $S_0>0$, $\rho<0$, $\mu\in\bbR$, $\lambda>0$, $H$ is a subordinator without drift, and $W$ is a $1$-dimensional standard Brownian motion.
Remark that we exclude the case of $\rho=0$, that is, the asset price process $S$ is continuous, since such a case is not interesting for our purposes.
Here $\Sigma$ is the volatility process defined via its squared process $\Sigma^2$, which is the solution to the stochastic differential equation
\[
d\Sigma_t^2 = -\lambda\Sigma_t^2dt+dH_{\lambda t}, \ \ \ t\in[0,T]
\]
with $\Sigma_0^2>0$, that is, $\Sigma^2$ is an Ornstein-Uhlenbeck process driven by the subordinator $H_\lambda$.
Note that $S$ is defined on some filtered probability space $(\Omega,\calF,(\calF_t)_{0\leq t\leq T},\bbP)$ with the usual condition,
where $(\calF_t)_{0\leq t\leq T}$ is the filtration generated by $W$ and $H_\lambda$.
Moreover, $X$ denotes the log price process $\log S$, that is, $X_t:=\log S_t$ for $t\in[0,T]$.
For more details on the BNS model, see also Nicolato and Venardos \cite{NV} and Schoutens \cite{Scho}.

For later use, we introduce some properties of $\Sigma$: Firstly, we have
\begin{equation}\label{sigma1}
\Sigma_t^2 = e^{-\lambda t}\Sigma^2_0+\int_0^te^{-\lambda(t-u)}dH_{\lambda u} \geq e^{-\lambda T}\Sigma^2_0
\end{equation}
for any $t\in[0,T]$, that is, $\Sigma$ is bounded from below.
Next, the integrated squared volatility is represented as
\begin{equation}\label{sigma2}
\int_t^T\Sigma_u^2du=\epsilon(T-t)\Sigma^2_t+\int_t^T\epsilon(T-u)dH_{\lambda u}
\end{equation}
for any $t\in[0,T]$, where
\[
\epsilon(t):=\frac{1-e^{-\lambda t}}{\lambda}.
\]
Further, we denote by $N$ the Poisson random measure of $H_\lambda$.
Hence, we have
\[
H_{\lambda t}=\int_0^\infty zN([0,t],dz), \ \ \ t\in[0,T].
\]
Letting $\nu$ be the L\'evy measure of $H_\lambda$, we find that
\[
\tN(dt,dz):=N(dt,dz)-\nu(dz)dt
\]
is the compensated Poisson random measure.
Note that $\nu$ is a $\sigma$-finite measure on $(0,\infty)$ satisfying
\begin{equation}\label{nu-cond1}
\int_0^\infty(z\wedge1)\nu(dz)<\infty
\end{equation}
by Proposition 3.10 of Cont and Tankov \cite{CT}.

Now, we introduce our standing assumptions as follows:

\begin{ass}\label{ass0}
\begin{enumerate}
\item $\displaystyle{\mu=\int_0^\infty(1-e^{\rho z})\nu(dz)}$.
\item $\displaystyle{\int_1^\infty e^{2\epsilon(T)z}\nu(dz)<\infty}$.
\end{enumerate}
\end{ass}

\noindent
Note that the finiteness of the right hand side of the condition 1 is ensured by (\ref{nu-cond1}).
The above condition 1 implies that the discounted asset price process $\whS_t:=e^{-rt}S_t$ becomes a local martingale.
On the other hand, the condition 2 ensures that
\[
\int_0^\infty z^2\nu(dz)<\infty,
\]
which yields $\bbE[H_{\lambda T}^2]<\infty$ by Proposition 3.13 of \cite{CT}.
In addition,
\begin{equation}\label{eq-S2}
\bbE\l[\sup_{t\in[0,T]}S^2_t\r]<\infty
\end{equation}
holds under the condition 2 from the view of Subsection 2.3 of Arai and Suzuki \cite{AS}.
Thus, $\whS$ is a square-integrable martingale under Assumption \ref{ass0}.
Moreover, the condition 2, together with (\ref{nu-cond1}), implies
\begin{equation}\label{nu-cond2}
\int_0^\infty z\nu(dz)<\infty.
\end{equation}

\begin{ex}\label{ex1}
We introduce two important examples of the squared volatility process $\Sigma^2$.
\begin{enumerate}
\item The first one is the case where $\Sigma^2$ follows an IG-OU process. The corresponding L\'evy measure $\nu$ is given by
      \[
      \nu(dz)=\frac{\lambda a}{2\sqrt{2\pi}}z^{-\frac{3}{2}}(1+b^2z)\exp\l\{-\frac{1}{2}b^2z\r\}dz, \ \ \ z\in(0,\infty),
      \]
      where $a>0$ and $b>0$.
      Note that this is a representative example of the BNS model with infinite active jumps, that is, $\nu((0,\infty))=\infty$.
      In this case, the invariant distribution of $\Sigma^2$ follows an inverse-Gaussian distribution with parameters $a>0$ and $b>0$.
      Note that the condition 2 of Assumption \ref{ass0} is satisfied whenever $\frac{b^2}{2}>2\epsilon(T)$
\item The second example is the gamma-OU case. In this case, $\nu$ is described as
      \[
      \nu(dz)=\lambda abe^{-bz}dz, \ \ \ z\in(0,\infty),
      \]
      and the invariant distribution of $\Sigma^2$ is given by a gamma distribution with parameters $a>0$ and $b>0$.
      If $b>2\epsilon(T)$, then the condition 2 of Assumption \ref{ass0} is satisfied.
\end{enumerate}
\end{ex}

Lastly, but not least, we introduce additional assumptions on the L\'evy measure $\nu$ as follows:

\begin{ass}\label{ass1}
The L\'evy measure $\nu$ is of the form $\nu(dz)=f(z)dz$ for $z>0$,
where $f(z)$ is a decreasing function on $(0,\infty)$, and there exists constants $C^\nu_0>0$ and $C^\nu_1>0$such that
\begin{equation}\label{ass1-1}
f(z) \leq C^\nu_0\gamma(z)e^{-C^\nu_1z}
\end{equation}
for any $z>0$, where $\gamma(z) := z^{-\frac{3}{2}}\vee z^{-\frac{1}{2}}$.
\end{ass}

\begin{rem}
Both examples introduced in Example \ref{ex1}: the IG-OU and the gamma-OU cases satisfy Assumption \ref{ass1}.
\end{rem}

\subsection{Black-Scholes formula}
We consider the Black-Scholes model in this subsection.
Let $\sigma>0$ be the volatility. For the call option with strike price $K>0$ and maturity $T>0$,
its price at time $t\in[0,T)$ with the log asset price $x\in\bbR$ is well-known as the Black-Scholes formula.
We treat it as a function on $\sigma^2$ as well as $t$ and $x$, and denote it by $BS(t,x,\sigma^2)$.
Recall that the function $BS$ is described as
\begin{equation}\label{def-BS}
BS(t,x,\sigma^2):=e^x\Phi(d^+)-Ke^{-r\tau_t}\Phi(d^-), \ \ \ t\in[0,T), x\in\bbR, \sigma>0.
\end{equation}
Here $\tau_t:=T-t$, $\Phi$ is the cumulative distribution function of the standard normal distribution, and
\[
d^\pm:=\frac{x-\log K+r\tau_t}{\sigma\sqrt{\tau_t}}\pm\frac{\sigma\sqrt{\tau_t}}{2},
\]
where double sign corresponds.
For later use, we define additionally
\[
d^\pm_{\rho z}:=d^\pm+\frac{\rho z}{\sigma\sqrt{\tau_t}}.
\]
for $z>0$.
Remark that the time parameter $t$ included in $d^\pm$ and $d^\pm_{\rho z}$ might be replaced with $u$ according to the situation.
In addition, since we have
\[
\lim_{t\to T}BS(t,x,\sigma^2)=(e^x-K)^+,
\]
the domain of the function $BS$ can be extended to $[0,T]\times\bbR\times(0,\infty)$, and we may define
\[
BS(T,x,\sigma^2):=(e^x-K)^+.
\]
Lastly, we define the following operators for $\bbR$-valued functions $f(t,x,\sigma^2)$, $t\in[0,T)$, $x\in\bbR$, $\sigma>0$:
\[
\Delta^{a,b}f(t,x,\sigma^2):=f(t,x+a,\sigma^2+b)-f(t,x,\sigma^2), \ \ \ a,b\in\bbR,
\]
\[
\calL^zf(t,x,\sigma^2):=\Delta^{\rho z,0}f(t,x,\sigma^2)+\partial_xf(t,x,\sigma^2)(1-e^{\rho z}), \ \ \ z>0,
\]
and
\[
\olcalL f(t,x,\sigma^2):=\int_0^\infty\calL^zf(t,x,\sigma^2)\nu(dz).
\]

\subsection{Decomposition formula}
For the BNS model introduced in Subsection 2.1, the price of the call option with strike price $K>0$ and maturity $T>0$ at time $t\in[0,T]$ is given as
\begin{equation}\label{eq-price}
V_t:=e^{-r\tau_t}\bbE[BS(T,X_T,\Sigma^2_T)|X_t,\Sigma^2_t],
\end{equation}
where $X_t=\log S_t$ for $t\in[0,T]$, since the discounted asset price process $\whS$ is a square-integrable martingale under Assumption \ref{ass0}.
The following theorem gives a decomposition expression of $V_t$, which will play important role for the development of approximations for (\ref{eq-price}).

\begin{thm}[Theorem 3.1 of Arai \cite{A}]\label{thm0}
Under Assumption \ref{ass0}, we have, for $t\in[0,T]$,
\begin{equation}\label{eq-thm0}
V_t = BS(t,X_t,\Sigma^2_t)+\tau_t\olcalL BS(t,X_t,\Sigma^2_t)+I_1+I_2+I_3+I_4+I_5.
\end{equation}
Here, $I_1,\dots,I_5$ are defined as follows:
\begin{align*}
I_1 &:= \bbE\l[\int_t^Te^{-r(u-t)}\partial_{\sigma^2}BS(u,X_u,\Sigma^2_u)(-\lambda\Sigma^2_u)du\Big|X_t,\Sigma^2_t\r], \hspace{90mm} \\
I_2 &:= \bbE\l[\int_t^Te^{-r(u-t)}\int_0^\infty\l(\Delta^{\rho z,z}-\Delta^{\rho z,0}\r)BS(u,X_u,\Sigma^2_u)\nu(dz)du\Big|X_t,\Sigma^2_t\r], \\
I_3 &:= \bbE\l[\int_t^Te^{-r(u-t)}\tau_u\partial_x\olcalL BS(u,X_u,\Sigma^2_u)\mu du\Big|X_t,\Sigma^2_t\r], \\
I_4 &:= \bbE\l[\int_t^Te^{-r(u-t)}\tau_u\partial_{\sigma^2}\olcalL BS(u,X_u,\Sigma^2_u)(-\lambda\Sigma^2_u)du\Big|X_t,\Sigma^2_t\r], \\
\mbox{ and } \\
I_5 &:= \bbE\Bigg[\int_t^Te^{-r(u-t)}\tau_u\int_0^\infty\Delta^{\rho z,z}\olcalL BS(u,X_u,\Sigma^2_u)\nu(dz)du\Big|X_t,\Sigma^2_t\Bigg],
\end{align*}
where $\tau_u:=T-u$.
\end{thm}

%
%
\setcounter{equation}{0}
\section{Main results}
In this section, we present approximate option pricing formulas, and illustrate some numerical results.
Henceforth, we assume Assumptions \ref{ass0} and \ref{ass1}.

\subsection{Approximation formulas}
For the call option price at time $t\in[0,T]$ with strike price $K>0$ for the BNS model,
we derive numerically tractable approximations from the decomposition formula (\ref{eq-thm0}).
First of all, the first term $BS(t,X_t,\Sigma^2_t)$ in (\ref{eq-thm0}) is treated as the primal term in our approximations, since it is easily computable.
On the other hand, all the remaining terms are not numerically tractable.
Thus, we need to exclude or replace them from our approximations.
In fact, we shall show that $I_1,\dots,I_5$ converge to 0 as the time to maturity goes to 0 with order $\frac{3}{2}$ in Proposition \ref{prop1} below.
This fact allows us to exclude $I_1,\dots,I_5$ from our approximations.
Hence, all what we have to do is to approximate the second term $\tau_t\olcalL BS(t,X_t,\Sigma^2_t)$.
To this end, we suggest two correction terms to $\tau_t\olcalL BS(t,X_t,\Sigma^2_t)$, both of which are computable.
One is givens as
\begin{equation}\label{eq-correction1}
\tau_t\int_{Z^0\vee\olZ}^\infty\l(Ke^{-r\tau_t}\Phi(D^-)-e^{X_t+\rho z}\Phi(D^+)\r)\nu(dz),
\end{equation}
where $\Phi$ is the cumulative distribution function of the standard normal distribution,
\[
Z^0:=\frac{X_t-\log K+r\tau_t}{|\rho|}, \ \ \ \olZ:=\frac{2\Sigma_t^2}{|\rho|} \ \ \
\mbox{ and } \ \ \ D^\pm:=\frac{X_t-\log K+r\tau_t}{\Sigma_t\sqrt{\tau_t}}\pm\frac{\Sigma_t\sqrt{\tau_t}}{2}.
\]
(\ref{eq-correction1}) is available for ITM and around ATM options, more precisely the case where $X_t-\log K>-2\Sigma^2_t$.
Another correction term is given as follows:
\begin{equation}\label{eq-correction2}
\tau_t\int_{Z_0}^\infty\l(Ke^{-r\tau_t}-e^{X_t+\rho z}\r)\nu(dz),
\end{equation}
which is defined for ITM options alone.

\begin{rem}\label{rem1}
We can give a financial interpretation for the correction term (\ref{eq-correction2}).
Consider an ITM option at time $t$.
Even if $\tau_t$ is small, the ITM option may change in a moment into an out-of-the-money (OTM) one due to a big jump in the asset price.
Thus, we can interpret that the integrand in the correction term (\ref{eq-correction2}) eliminates the payoff of the ITM option when a big jump occurs, roughly speaking.
In addition, since the probability that such a big jump occurs is nearly equal to $\tau_t\nu([Z^0,\infty))$, the correction term is multiplied by $\tau_t$.
In contrast, since positive jumps never occur in the asset price process, we do not need to take into account the reverse changes, that is,
the changes of an OTM option into an ITM one. Hence, the correction term is available for ITM options alone.

On the other hand, (\ref{eq-correction1}) is available for not only ITM but also around ATM options.
The integrand in (\ref{eq-correction1}) is close to that in (\ref{eq-correction2}) when $X_t-\log K>0$ since $D^\pm$ tends to $\infty$ as $\tau_t\to0$,
while $D^\pm$ tends to $-\infty$ as $\tau_t\to0$ when $X_t-\log K<0$, that is, (\ref{eq-correction1}) is close to 0 for OTM options.
\end{rem}

Theorems \ref{thm1} and \ref{thm2} provide approximation formulas using the above correction terms (\ref{eq-correction1}) and (\ref{eq-correction2}), respectively, and
a hybrid version is introduced in Corollary \ref{cor1}.

\begin{thm}\label{thm1}
Let
\[
\tV^1_t:=BS(t,X_t,\Sigma^2_t)+\tau_t\int_{Z^0\vee\olZ}^\infty\l(Ke^{-r\tau_t}\Phi(D^-)-e^{X_t+\rho z}\Phi(D^+)\r)\nu(dz).
\]
When $X_t-\log K>-2\Sigma_t^2$, there is a constant $C>0$ such that
\[
\l|V_t-\tV^1_t\r|\leq C\l\{\tau_t^{\frac{3}{2}}{\bf 1}_{\{X_t-\log K\geq2\Sigma_t^2\}}+\tau_t{\bf 1}_{\{|X_t-\log K|<2\Sigma_t^2\}}\r\}
\]
for any $t\in[0,T]$, where $C>0$ is depending on $X_t$ and $\Sigma_t$, and a nondecreasing function of $T$.
\end{thm}

\begin{thm}\label{thm2}
Suppose $X_t-\log K\geq2\Sigma_t^2$, and denote
\[
\tV^2_t:=BS(t,X_t,\Sigma^2_t)+\tau_t\int_{Z^0}^\infty\l(Ke^{-r\tau_t}-e^{X_t+\rho z}\r)\nu(dz).
\]
There is a constant $C>0$ such that
\[
\l|V_t-\tV^2_t\r|\leq C\tau_t^{\frac{3}{2}}
\]
for any $t\in[0,T]$, where $C>0$ is depending on $X_t$ and $\Sigma_t$, and a nondecreasing function of $T$.
\end{thm}

\begin{cor}\label{cor1}
Suppose  $X_t-\log K>-2\Sigma_t^2$, and denote
\begin{align*}
\tV^3_t &:= BS(t,X_t,\Sigma^2_t)+\tau_t\int_{Z^0}^\infty\l(Ke^{-r\tau_t}-e^{X_t+\rho z}\r)\nu(dz){\bf 1}_{\{X_t-\log K\geq2\Sigma_t^2\}} \\
        &\hspace{5mm} +\tau_t\int_{Z^0\vee\olZ}^\infty\l(Ke^{-r\tau_t}\Phi(D^-)-e^{X_t+\rho z}\Phi(D^+)\r)\nu(dz){\bf 1}_{\{|X_t-\log K|<2\Sigma_t^2\}}.
\end{align*}
There is a constant $C>0$ such that
\[
\l|V_t-\tV^3_t\r|\leq C\tau_t
\]
for any $t\in[0,T]$, where $C>0$ is depending on $X_t$ and $\Sigma_t$, and a nondecreasing function of $T$.
\end{cor}

\begin{rem}\label{rem2}
Any approximation introduced in this section is not available when the option is deep OTM, that is, the case of $X_t-\log K\leq-2\Sigma_t^2$,
but this exclusion is not restrictive, since deep OTM option prices with short maturity are sufficiently near to zero.

Moreover, in spite that the correction term (\ref{eq-correction2}) is defined for all ITM options,
Theorem \ref{thm2} excludes the case where $2\Sigma_t^2>X_t-\log K>0$, that is, the option is ITM, but near to ATM.
It is because $Z^0$ takes a small positive value in such a case, hence the integration in (\ref{eq-correction2}) takes a huge value since $\nu(dz)$ diverges as $z\to0$.
In fact, as shown in Figure \ref{fig1} below, the difference between $\tV^2_t$ and the true option price $V_t$ increases rapidly when $X_t-\log K$ goes down to 0, that is, $K\to468.44$.
Therefore, we restrict the value of $X_t-\log K$ to be greater than $2\Sigma_t^2$ in Theorem \ref{thm2}.
On the other hand, Theorem \ref{thm1} is useful for around ATM options, too, but the rate of convergence for around ATM options is changed from $\tau_t^{\frac{3}{2}}$ into $\tau_t$.

\begin{figure}[H]
\centering\includegraphics[width=70mm]{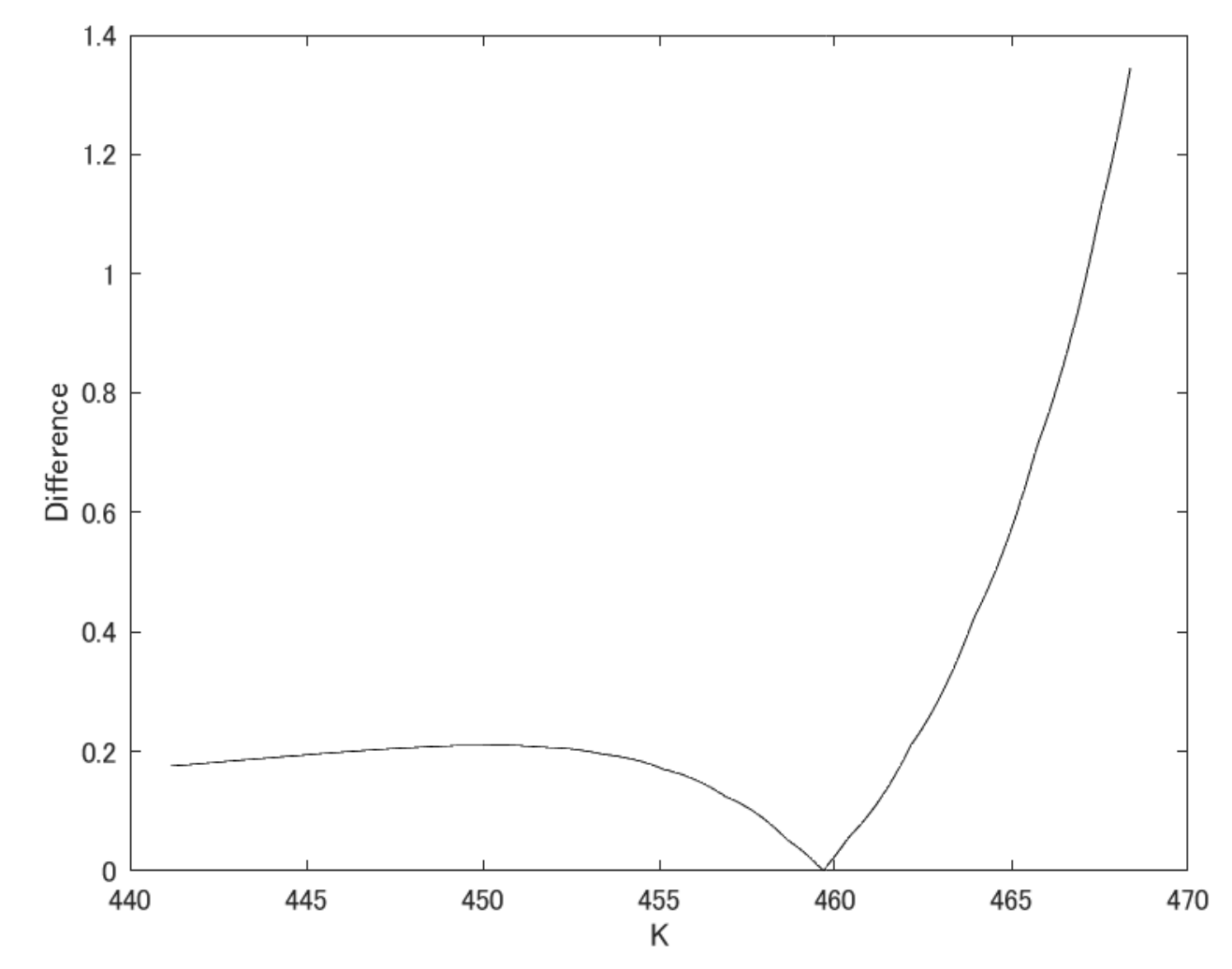}
\vspace{-3mm}\caption{The difference $|\tV^2_t-V_t|$ versus strike price $K$ for ITM options
when $t=0$ and $T=0.0833$ with parameter set $NV$ introduced in Table \ref{tab1}.
Note that the option is ATM when $K=468.44$. For more details on numerical method, see Subsection 3.2.}\label{fig1}
\end{figure}
\end{rem}

\subsection{Numerical experiments}
We implement some numerical experiments in order to examine the effectiveness of approximation formulas $\tV^1$ and $\tV^3$ introduced in Theorem \ref{thm1}
and Corollary \ref{cor1}, respectively.
Note that $\tV^2$ in Theorem \ref{thm2} is excluded from our numerical experiments, since it is a part of $\tV^3$.
More specifically, we treat the IG-OU case introduced in Example \ref{ex1},
and compute the relative errors of $\tV^1_t$ and $\tV^3_t$ to the true option price $V_t$ defined in (\ref{eq-price}), where the relative error of $U_t$ to $V_t$ is defined as
\[
\frac{|U_t-V_t|}{V_t}.
\]
For comparison, we shall compute simultaneously the relative errors of $BS(t,X_t,\Sigma^2_t)$.

Now, we explain how to compute the correction terms (\ref{eq-correction1}) and (\ref{eq-correction2}).
Recall that the L\'evy measure of the IG-OU case is given as
\[
\nu(dz)=\frac{\lambda a}{2\sqrt{2\pi}}z^{-\frac{3}{2}}(1+b^2z)\exp\l\{-\frac{1}{2}b^2z\r\}{\bf 1}_{(0,\infty)}(z)dz,
\]
where $a>0$ and $b>0$.
To simplify the notation, we denote
\[
c_0:=\frac{\lambda a}{2\sqrt{2\pi}}, \ \ \ c_1:=\frac{\lambda ab^2}{2\sqrt{2\pi}} \ \ \ \mbox{and} \ \ \ c_2:=\frac{b^2}{2}.
\]
For $\alpha$, $\beta>0$, we define a function $\Gamma(\alpha,\beta)$ as
\[
\Gamma(\alpha,\beta):=\int_\beta^\infty e^{-z}z^{\alpha-1}dz,
\]
which is called the upper incomplete gamma function.
Denoting
\[
\Gamma^1(c,\beta):=\int_\beta^\infty e^{-cz}z^{-\frac{1}{2}}dz=\frac{1}{\sqrt{c}}\Gamma\l(\frac{1}{2},\beta c\r)
\]
and
\[
\Gamma^3(c,\beta):=\int_\beta^\infty e^{-cz}z^{-\frac{3}{2}}dz=\frac{2e^{-\beta c}}{\sqrt{\beta}}-2\sqrt{c}\Gamma\l(\frac{1}{2},\beta c\r)
\]
for $c>0$ and $\beta>0$, we have
\[
\nu([u,\infty))=c_0\Gamma^3(c_2,u)+c_1\Gamma^1(c_2,u)
\]
for any $u>0$. Thus, denoting
\[
z^0 := \frac{x-\log K+r\tau_t}{|\rho|} \ \ \ \mbox{ and } \ \ \ \olz := \frac{2\sigma^2}{|\rho|},
\]
we can compute (\ref{eq-correction1}) and (\ref{eq-correction2}) as follows:
\begin{align*}
&\int_{z^0\vee\olz}^\infty\l(Ke^{-r\tau_t}\Phi(d^-)-e^{x+\rho z}\Phi(d^+)\r)\nu(dz) \\
&= Ke^{-r\tau_t}\Phi(d^-)\nu([z^0\vee\olz,\infty))-e^x\Phi(d^+)\int_{z^0\vee\olz}^\infty e^{\rho z}\nu(dz) \\
&= Ke^{-r\tau_t}\Phi(d^-)\l\{c_0\Gamma^3(c_2,z^0\vee\olz)+c_1\Gamma^1(c_2,z^0\vee\olz)\r\} \\
&\hspace{5mm} -e^x\Phi(d^+)\l\{c_0\Gamma^3(c_2+|\rho|,z^0\vee\olz)+c_1\Gamma^1(c_2+|\rho|,z^0\vee\olz)\r\},
\end{align*}
and, when $x-\log K\geq2\sigma^2$,
\begin{align*}
&\int_{z^0}^\infty(Ke^{-r\tau_t}-e^{x+\rho z})^+\nu(dz) \\
&= Ke^{-r\tau_t}\nu([z^0,\infty))-e^x\int_{z^0}^\infty e^{\rho z}\nu(dz) \\
&= Ke^{-r\tau_t}\l\{c_0\Gamma^3(c_2,z^0)+c_1\Gamma^1(c_2,z^0)\r\}-e^x\l\{c_0\Gamma^3(c_2+|\rho|,z^0)+c_1\Gamma^1(c_2+|\rho|,z^0)\r\}.
\end{align*}

Next, we introduce the numerical experiments implemented here.
The following two parameter sets for the IG-OU case will be used:

\begin{table}[htb]\begin{center}
\caption{Parameter sets used in numerical experiments}\label{tab1}
\begin{tabular}{cccccccc} \hline \vspace{-3.5mm} \\
      & $\rho$  & $\lambda$ & $a$    & $b$    & $\Sigma_t^2$ & $S_t(=e^{X_t})$ & $r$    \\ \hline \hline
$NV$  & -4.7039 & 2.4958    & 0.0872 & 11.98  & 0.0041       & 468.44          & 0.0319 \\ \hline
$Sch$ & -0.1926 & 0.0636    & 6.2410 & 0.7995 & 0.0156       & 1124.47         & 0.007  \\ \hline
\end{tabular}\end{center}\end{table}

\noindent
The parameter sets $NV$ and $Sch$ come from Table 5.1 of \cite{NV} and Table 7.1 of \cite{Scho}, respectively.
Both of them meet Assumption \ref{ass0} and are estimated from S\&P 500 index option prices data on November 2, 1993, and on April 18, 2002, respectively.
Note that $t$ is fixed to 0 throughout.
We implement three types of numerical experiments.
First, we compute the relative errors of $BS(0,X_0,\Sigma^2_0)$, $\tV^1_0$ and $\tV^3_0$ for ITM and around ATM call options when $T$ is fixed to 0.0833($\fallingdotseq$1/12),
where the values of $V_0$ are computed by the fast Fourier transform-based numerical scheme developed in Section 6 of Arai et al. \cite{AIS-BNS}.
Figure \ref{fig2} illustrates the result of this experiment and indicates that $\tV^1_0$ and $\tV^3_0$ give very nice approximations of $V_0$ for any ITM and around ATM options.
Roughly speaking, the performance of $\tV^3_0$ is better than that of $\tV^1_0$ for deep ITM options.
In Figure \ref{fig3}, varying $T$ from 0.01 to 0.4, the relative errors of $BS(0,X_0,\Sigma^2_0)$ and $\tV^1_0$ for the ATM option are computed.
Note that $\tV^3$ is excluded since it takes the same values as $\tV^1$ when the option is ATM.
On the other hand, the relative errors for an ITM option are computed in Figure \ref{fig4} by fixing $K$ to $e^{X_0-2\Sigma^2_0}-0.02e^{X_0}$.
In this case, the values of $\tV^1_0$ and $\tV^3_0$ are not overlapped, thus, the relative errors of $\tV^3_0$ are computed as well as $BS(0,X_0,\Sigma^2_0)$ and $\tV^1_0$.
From Figures \ref{fig3} and \ref{fig4}, we can say that the relative errors of $\tV^1_0$ and $\tV^3_0$ are much smaller than those of $BS(0,X_0,\Sigma^2_0)$,
but the bigger the time to maturity is, the worse the effectiveness of $\tV^1_0$ and $\tV^3_0$ is.
Thus, when the time to maturity is away from 0, the relative errors of $\tV^1_0$ and $\tV^3_0$ are not sufficiently small.
As for the comparison of performance between $\tV^1_0$ and $\tV^3_0$ in Figure \ref{fig4}, $\tV^3_0$ is better in Panel (NV), but it is reversed in Panel (Sch).
In summary, the approximations $\tV^1$ and $\tV^3$ are effective for ITM and around ATM options with short maturity, regardless of the choice of parameter sets.

\renewcommand{\figurename}{Panel}\renewcommand{\thefigure}{(NV)}
\begin{figure}[H]
\begin{minipage}{0.5\hsize}
    \includegraphics[width=70mm]{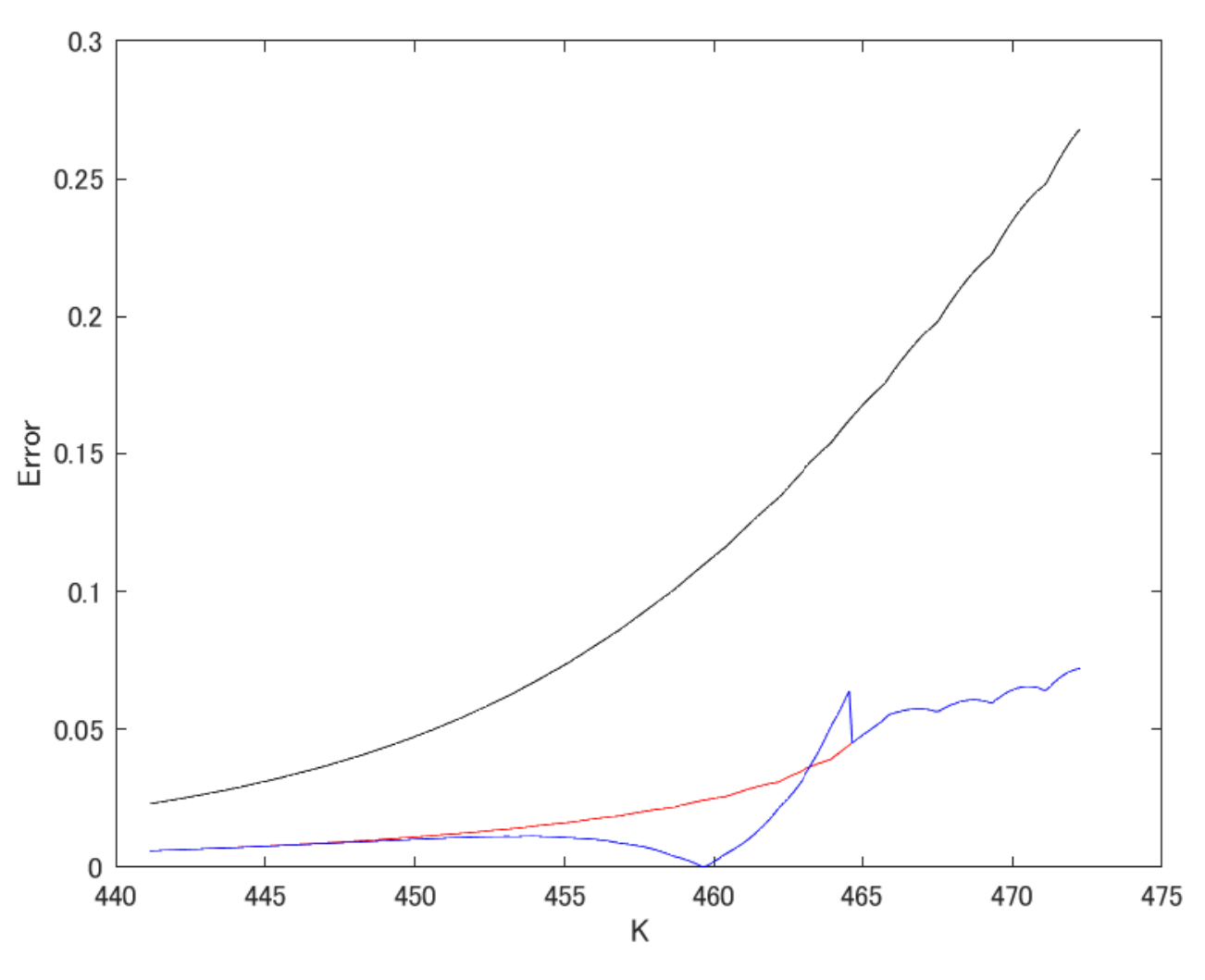}\vspace{-2mm}\caption{}\label{fig2NV}
\end{minipage}
\renewcommand{\thefigure}{(Sch)}
\begin{minipage}{0.5\hsize}
    \includegraphics[width=70mm]{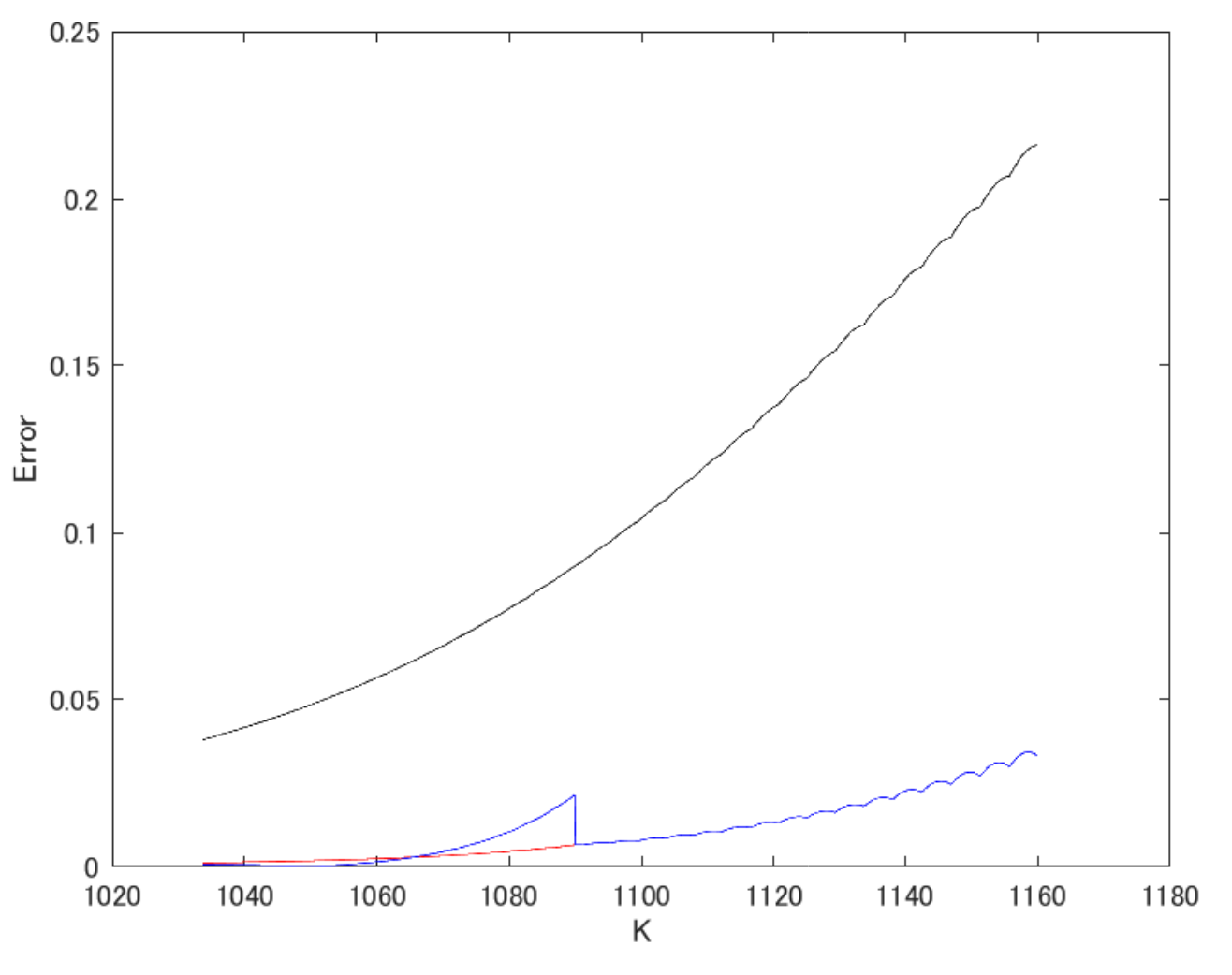}\vspace{-2mm}\caption{}\label{fig2Scho}
\end{minipage}
\renewcommand{\figurename}{Figure}\setcounter{figure}{1}\renewcommand{\thefigure}{\arabic{figure}}
\vspace{-3mm}\caption{The relative errors of $BS(0,X_0,\Sigma^2_0)$, $\tV^1_0$ and $\tV^3_0$ versus strike price $K$ when $T=0.0833$
with parameter set $NV$ in Panel \ref{fig2NV} and $Sch$ in Panel \ref{fig2Scho}.
The black, red and blue curves represent the relative errors of $BS(0,X_0,\Sigma^2_0)$, $\tV^1_0$ and $\tV^3_0$, respectively.
The values of $K$ vary from $\exp\{X_0-2\Sigma^2_0\}-0.05e^{X_0}$ to $\exp\{X_0+2\Sigma^2_0\}$, that is, from ITM to OTM being near to ATM.
Note that the red and blue curves are overlapped when $K>\exp\{X_0-2\Sigma^2_0\}$, that is, $K>464.5870$ in Panel \ref{fig2NV} and $K>1089.9$ in Panel \ref{fig2Scho}, respectively.}
\label{fig2}
\end{figure}

\renewcommand{\figurename}{Panel}\renewcommand{\thefigure}{(NV)}
\begin{figure}[H]
\begin{minipage}{0.5\hsize}
    \includegraphics[width=70mm]{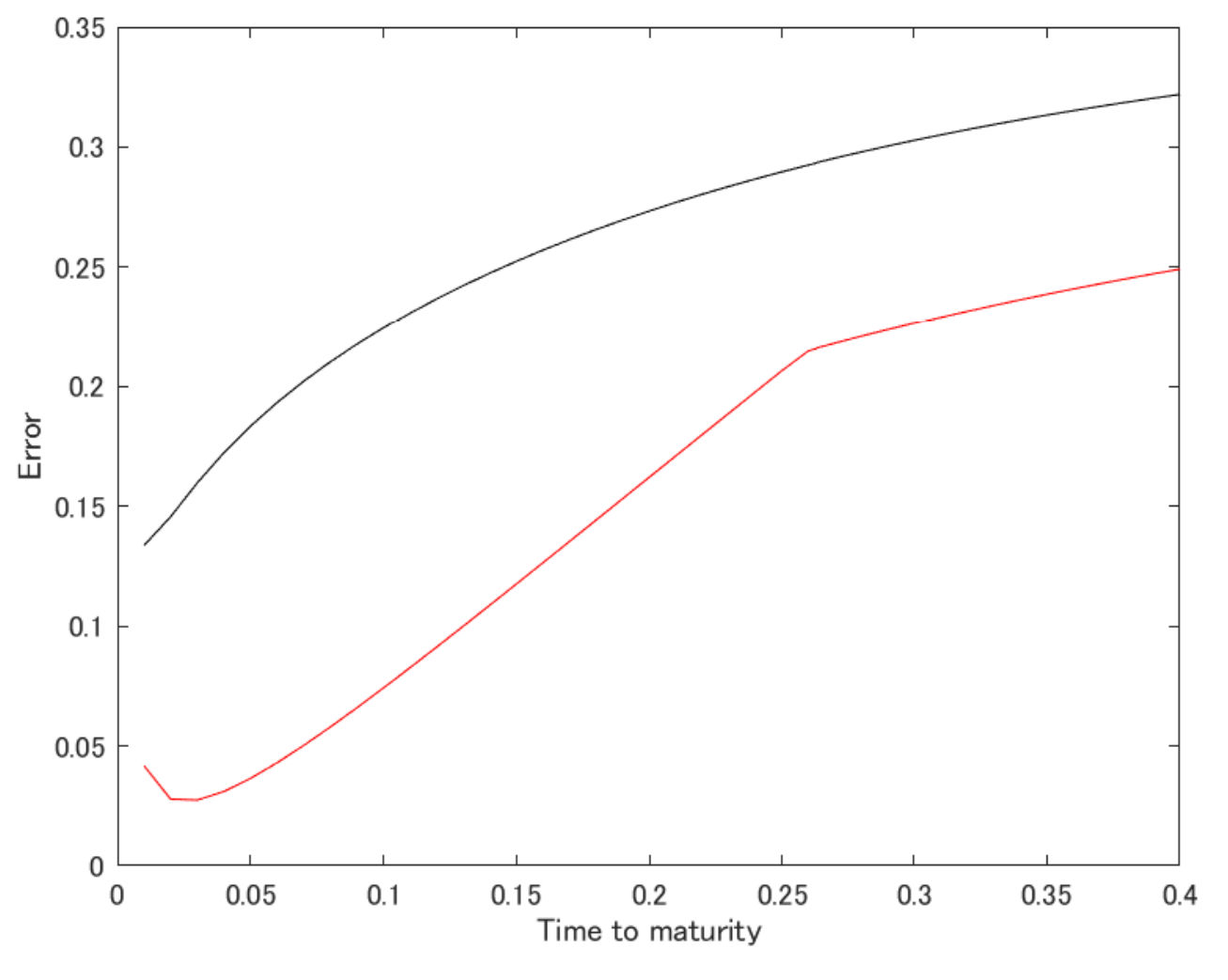}\vspace{-2mm}\caption{}\label{fig3NV}
\end{minipage}
\renewcommand{\thefigure}{(Sch)}
\begin{minipage}{0.5\hsize}
    \includegraphics[width=70mm]{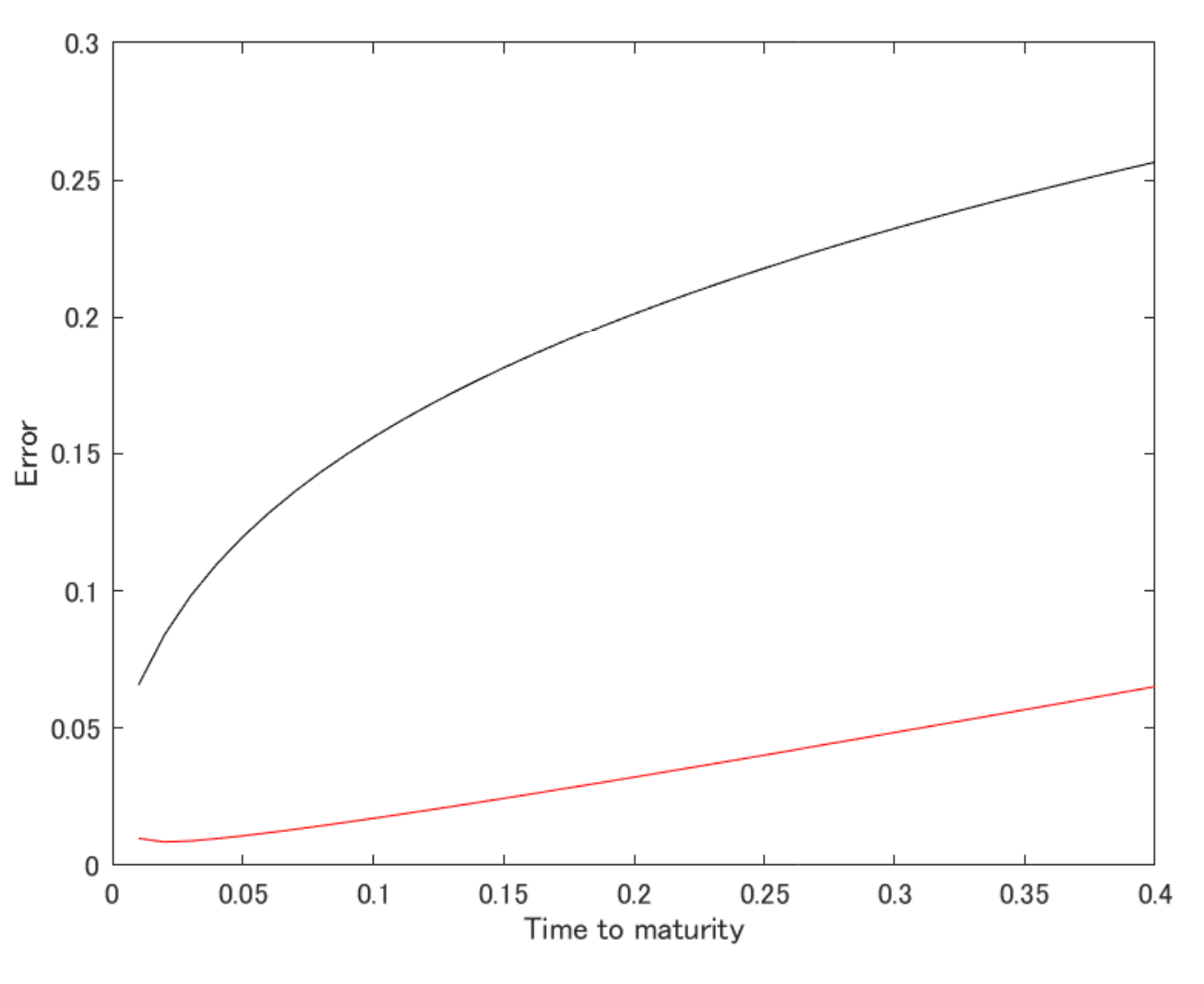}\vspace{-2mm}\caption{}\label{fig3Scho}
\end{minipage}
\renewcommand{\figurename}{Figure}\setcounter{figure}{2}\renewcommand{\thefigure}{\arabic{figure}}
\vspace{-3mm}\caption{The relative errors of $BS(0,X_0,\Sigma^2_0)$(black) and $\tV^1_0$(red) versus time to maturity from 0.01 to 0.4 when the option is ATM, that is,
$K=468.44$ in Panel \ref{fig3NV} and $K=1124.47$ in Panel \ref{fig3Scho}, respectively.
The same parameter sets as Figure \ref{fig2} are used.}\label{fig3}
\end{figure}

\renewcommand{\figurename}{Panel}\renewcommand{\thefigure}{(NV)}
\begin{figure}[H]
\begin{minipage}{0.5\hsize}
    \includegraphics[width=70mm]{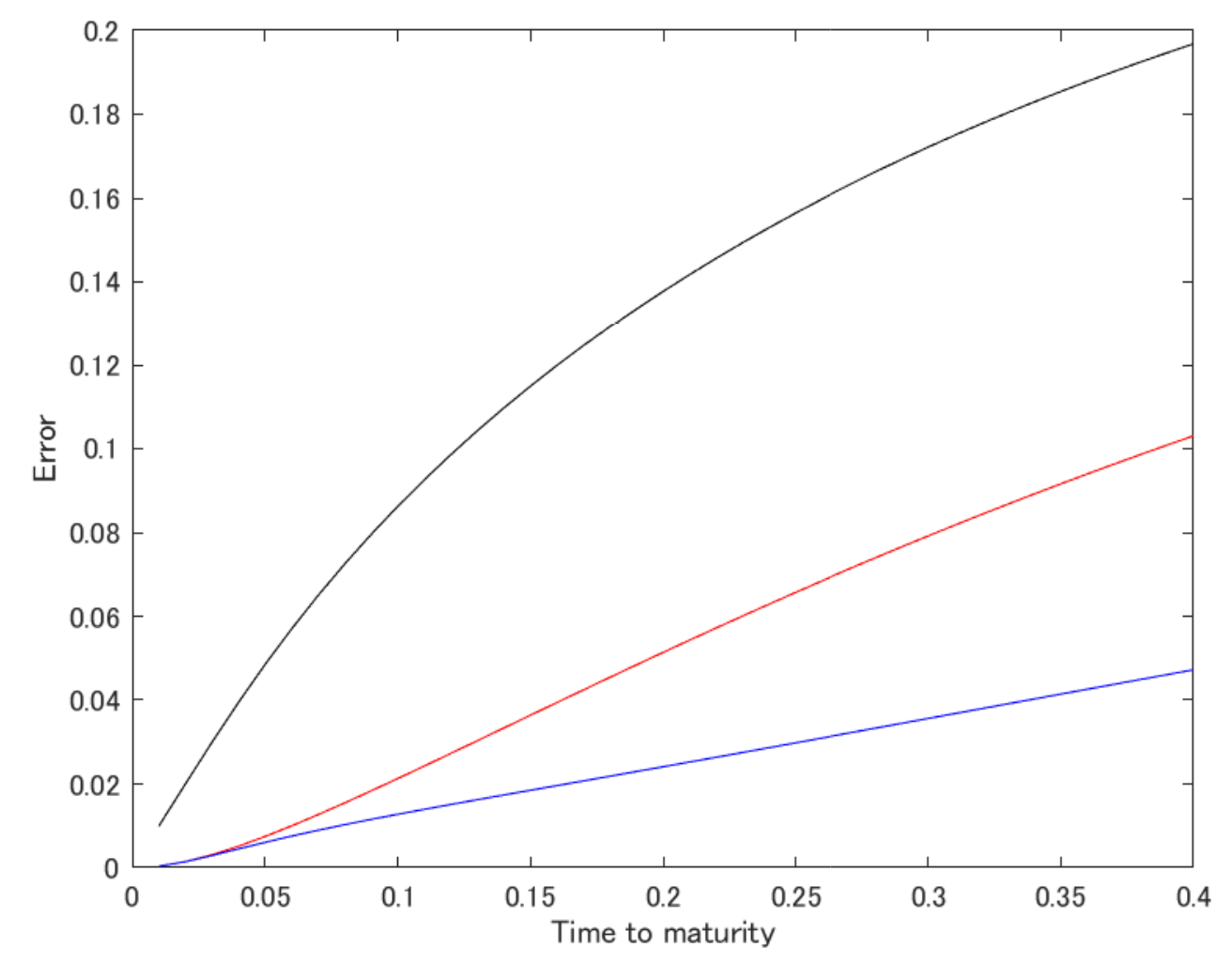}\vspace{-2mm}\caption{}\label{fig4NV}
\end{minipage}
\renewcommand{\thefigure}{(Sch)}
\begin{minipage}{0.5\hsize}
    \includegraphics[width=70mm]{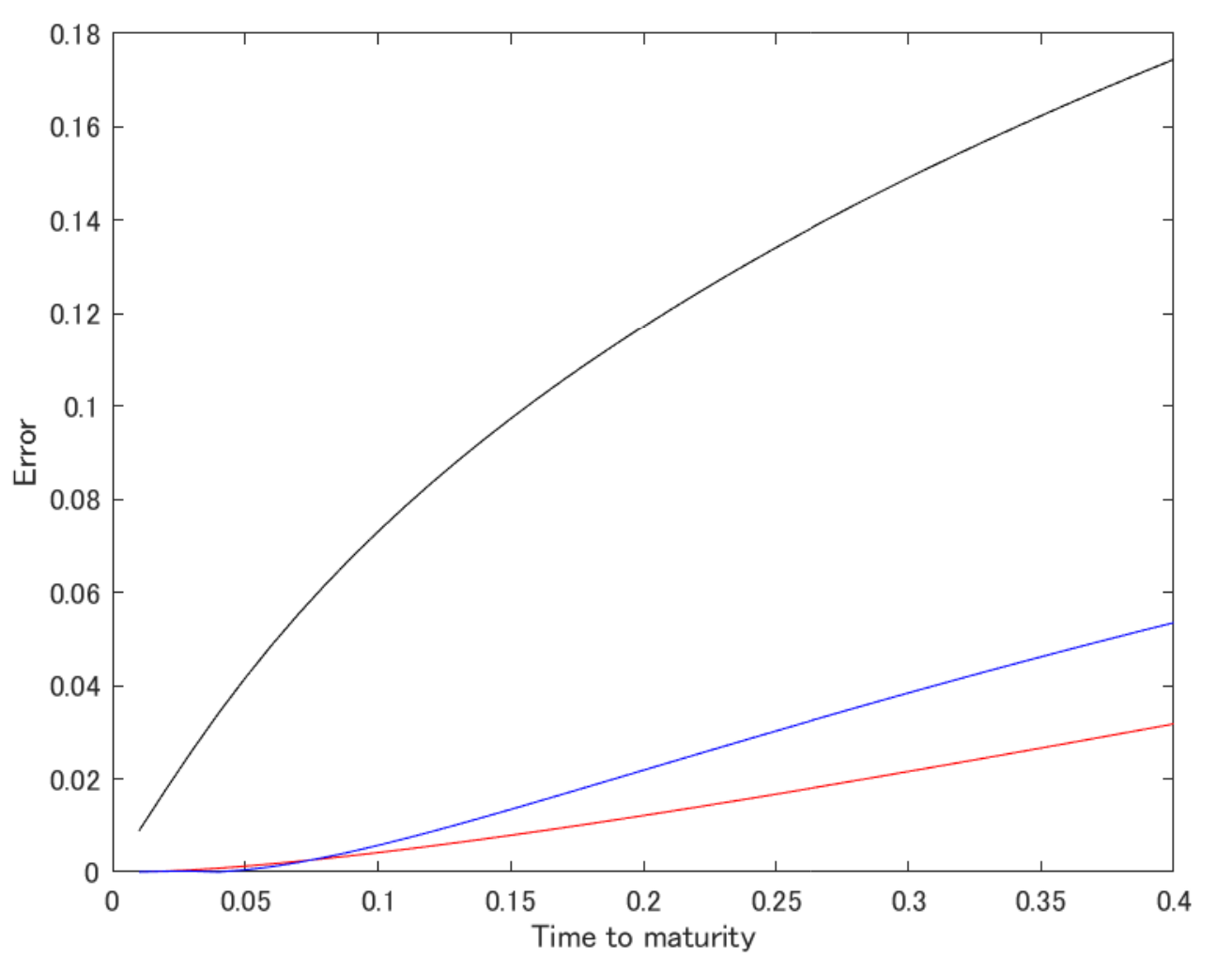}\vspace{-2mm}\caption{}\label{fig4Scho}
\end{minipage}
\renewcommand{\figurename}{Figure}\setcounter{figure}{3}\renewcommand{\thefigure}{\arabic{figure}}
\vspace{-3mm}\caption{The relative errors of $BS(0,X_0,\Sigma^2_0)$(black), $\tV^1_0$(red) and $\tV^3_0$(blue) versus time to maturity from 0.01 to 0.4
when the option is ITM, more precisely, $K$ is fixed to $\exp\{X_0-2\Sigma^2_0\}-0.02e^{X_0}$, that is, 455.2182 in Panel \ref{fig4NV} and 1067.4 in Panel \ref{fig4Scho}, respectively.}\label{fig4}
\end{figure}

%
%
\setcounter{equation}{0}
\section{Proofs}
We prove Theorems \ref{thm1} and \ref{thm2} in this section.
First of all, we give evaluations of $I_k$ for $k=1,\dots,5$ before proving the theorems.

\begin{prop}\label{prop1}
For $k=1,\dots,5$, there is a constant $C>0$ such that $|I_k|\leq C\tau_t^{\frac{3}{2}}$.
Remark that $C$ is depending on $X_t$, $\Sigma_t$ and $T$, and nondecreasing as a function of $T$.
\end{prop}

\proof
Remark that all constants $C>0$ appeared in this proof are depending on $X_t$ (or $x$), $\Sigma_t$ (or $\sigma$) and $T$, and nondecreasing as a function of $T$.
Firstly, we see $|I_1|\leq C\tau_t^{\frac{3}{2}}$ for some $C>0$.
Note that $e^x\phi(d^+)=Ke^{-r\tau_u}\phi(d^-)$ holds, where $\phi$ is the probability density function of the standard normal distribution.
Thus, (\ref{def-BS}) implies
\begin{equation}\label{eq-prop1-3}
\l|\partial_{\sigma^2}BS(u,x,\sigma^2)\r| = \frac{\sqrt{\tau_u}}{2\sigma}Ke^{-r\tau_u}\phi(d^-) \leq \frac{\sqrt{\tau_u}}{2\sigma}Ke^{-r\tau_u}\frac{1}{\sqrt{2\pi}}.
\end{equation}
This, together with (\ref{sigma1}) and (\ref{sigma2}), provides
\[
|I_1| \leq \frac{\lambda\sqrt{\tau_t}K}{2\sqrt{2\pi}e^{-\lambda T/2}\Sigma_0}\bbE\l[\int_t^T\Sigma^2_udu\Big|X_t,\Sigma^2_t\r]
      \leq \frac{\lambda\sqrt{\tau_t}K}{2\sqrt{2\pi}e^{-\lambda T/2}\Sigma_0}\l(\tau_t\Sigma^2_t+\frac{\tau_t^2}{2}\int_0^\infty z\nu(dz)\r)
      < C\tau_t^{\frac{3}{2}}
\]
for some $C>0$, since $\epsilon(t)\leq t$ holds for any $t\in[0,T]$.
Remark that $C$ is depending on $\Sigma_t$, and nondecreasing as a function of $T$.

Next, we show $|I_2|\leq C\tau_t^{\frac{3}{2}}$ for some $C>0$.
(\ref{eq-prop1-3}) implies that
\begin{align*}
|(\Delta^{\rho z,z}-\Delta^{\rho z,0})BS(u,x,\sigma^2)| 
&=    |BS(u,x_z,\sigma_z^2)-BS(u,x_z,\sigma^2)| = \l|\int_{\sigma^2}^{\sigma_z^2}\partial_{\sigma^2}BS(u,x_z,\tilde{\sigma}^2)d\tilde{\sigma}^2\r| \\
&\leq \sup_{\sigma^2\leq\tilde{\sigma}^2\leq\sigma_z^2}|\partial_{\sigma^2}BS(u,x_z,\tilde{\sigma}^2)|(\sigma_z^2-\sigma^2) \leq \frac{\sqrt{\tau_u}}{2\sqrt{2\pi}\sigma}Ke^{-r\tau_u}z,
\end{align*}
where $x_z:=x+\rho z$ and $\sigma_z^2:=\sigma^2+z$.
As a result, we obtain by (\ref{sigma1})
\[
|I_2| \leq \bbE\l[\int_t^T\int_0^\infty\frac{\sqrt{\tau_u}}{2\sqrt{2\pi}e^{-\lambda T/2}\Sigma_0}Ke^{-r\tau_u}z\nu(dz)du\Big|X_t,\Sigma^2_t\r] \leq C\tau_t^{\frac{3}{2}}
\]
for some $C>0$.

As for $I_3$, we have
\begin{equation}\label{eq-prop1-4}
|\partial_x\olcalL BS(u,x,\sigma^2)| \leq \frac{e^{x}}{\sqrt{2\pi}\sigma\sqrt{\tau_u}}\int_0^\infty\l(e^{\rho z}|\rho|z+1-e^{\rho z}\r)\nu(dz)
\end{equation}
by the proof of Lemma 4.4 in \cite{A}.
This implies that 
\begin{align*}
|I_3| &\leq \bbE\l[\int_t^T\frac{S_u\mu\sqrt{\tau_u}}{\sqrt{2\pi}e^{-\lambda T/2}\Sigma_0}\int_0^\infty\l(e^{\rho z}|\rho|z+1-e^{\rho z}\r)\nu(dz)du\Big|X_t,\Sigma^2_t\r] \\
      &\leq C\bbE\l[\sup_{t\in[0,T]}S_t\r]\tau_t^{\frac{3}{2}} \leq C\tau_t^{\frac{3}{2}}
\end{align*}
for some $C>0$ by (\ref{sigma1}) and (\ref{eq-S2}).
To see the same evaluation for $I_4$, (4.13) in \cite{A} provides
\begin{equation}\label{eq-prop1-5}
|\partial_{\sigma^2}\olcalL BS(u,x,\sigma^2)| \leq \frac{e^x}{2\sqrt{2\pi e}\sigma^2}\int_0^\infty\l(e^{\rho z}|\rho|z+1-e^{\rho z}\r)\nu(dz)
\end{equation}
where $|\phi^\prime(\vt)|\leq\frac{1}{\sqrt{2\pi e}}$ for any $\vt\in\bbR$.
Thus, by (\ref{eq-S2}), there exists a constant $C>0$ such that $|I_4|\leq C\tau_t^{\frac{3}{2}}$.

Next, we evaluate $|I_5|$.
From the views of (\ref{eq-prop1-4}) and (\ref{eq-prop1-5}), we have
\begin{align*}
& \l|\int_0^\infty\Delta^{\rho z,z}\olcalL BS(t,x,\sigma^2)\nu(dz)\r| \\
&\leq \int_0^\infty\l\{|\olcalL BS(t,x_z,\sigma_z^2)-\olcalL BS(t,x,\sigma_z^2)|+|\olcalL BS(t,x,\sigma_z^2)-\olcalL BS(t,x,\sigma^2)|\r\}\nu(dz) \\
&\leq \int_0^\infty\l\{|\rho|z\sup_{x_z\leq y\leq x}\l|\partial_x\olcalL BS(t,y,\sigma_z^2)\r|
      +z\sup_{\sigma^2\leq\tilde{\sigma}^2\leq\sigma_z^2}\l|\partial_{\sigma^2}\olcalL BS(t,x,\tilde{\sigma}^2)\r|\r\}\nu(dz) \\
&\leq Ce^x\int_0^\infty\l\{\frac{|\rho|z}{\sigma_z\sqrt{\tau_t}}+\frac{z}{\sigma^2}\r\}\nu(dz)
      \leq \frac{Ce^x}{\sigma\wedge\sigma^2}\l(\frac{1}{\sqrt{\tau_t}}+1\r),
\end{align*}
from which $|I_5| \leq C\tau_t^{\frac{3}{2}}$ follows for some $C>0$.
\fin

\subsection{Proof of Theorem \ref{thm1}}
From the view of Proposition \ref{prop1}, it is enough to show that there is a constant $C>0$ satisfying
\begin{align}\label{eq-thm1-1}
&\l|\olcalL BS(t,x,\sigma^2)-\int_{z^0\vee\olz}^\infty\l(Ke^{-r\tau_t}\Phi(d^-)-e^{x_z}\Phi(d^+)\r)\nu(dz)\r| \nonumber \\
&\leq C\l\{\tau_t^{\frac{1}{2}}{\bf 1}_{\{x-\log K\geq2\sigma^2\}}+{\bf 1}_{\{|x-\log K|<2\sigma^2\}}\r\}
\end{align}
for any $t\in[0,T)$, $\sigma^2\in[e^{-\lambda T}\Sigma^2_0,\infty)$ and $x>\log K-2\sigma^2$, where $x_z:=x+\rho z$ and
\[
z^0=\frac{x-\log K+r\tau_t}{|\rho|} \ \ \ \mbox{ and } \ \ \ \olz=\frac{2\sigma^2}{|\rho|}.
\]
Without loss of generality, we may assume that $\tau_t\in(0,1]$.
To this end, we decompose the left hand side of (\ref{eq-thm1-1}) into the following three terms:
\begin{align}\label{eq-thm1-2}
&\Bigg|\int_0^{\frac{z^0}{2}}\calL^z(t,x,\sigma^2)\nu(dz)\Bigg|+\Bigg|\int_{\frac{z^0}{2}}^{z^0\vee\olz}\calL^z(t,x,\sigma^2)\nu(dz)\Bigg| \nonumber \\
&\hspace{5mm} +\Bigg|\int_{z^0\vee\olz}^\infty\l\{\calL^z(t,x,\sigma^2)-\l(Ke^{-r\tau_t}\Phi(d^-)-e^{x_z}\Phi(d^+)\r)\r\}\nu(dz)\Bigg|.
\end{align}

\noindent{\it Step 1.} \ 
In this step, we treat the case where
\begin{equation}\label{eq-thm1-4}
x-\log K\geq2\sigma^2.
\end{equation}
We show that there is a constant $C>0$ such that each term of (\ref{eq-thm1-2}) is less than $C\sqrt{\tau_t}$ for any $\tau_t\in(0,1]$.
Note that $z^0\geq\olz$ holds in this case, and the constants $C>0$ appeared in this step are depending on $x$ and $\sigma$, but independent of $T$.

Firstly, we treat the first term of (\ref{eq-thm1-2}).
Note that (\ref{eq-thm1-4}) implies $0<d^-_{\rho z}$ for any $z\in(0,\frac{z^0}{2}]$.
We have then
\begin{align}\label{eq-thm1-6}
\frac{\phi(d^\pm_{\rho z})}{\sigma\sqrt{\tau_t}}
&=    \frac{1}{\sqrt{2\pi}\sigma\sqrt{\tau_t}}\exp\l\{-\frac{(x-\log K+r\tau_t+\rho z\pm\frac{\sigma^2}{2}\tau_t)^2}{2\sigma^2\tau_t}\r\} \nonumber \\
&\leq \frac{1}{\sqrt{2\pi}\sigma\sqrt{\tau_t}}\exp\l\{-\frac{\l(\frac{x-\log K+r\tau_t}{2}-\frac{\sigma^2}{2}\tau_t\r)^2}{2\sigma^2\tau_t}\r\} \nonumber \\
&\leq \frac{1}{\sqrt{2\pi}\sigma\sqrt{\tau_t}}\exp\l\{-\frac{\l(\sigma^2-\frac{\sigma^2}{2}\tau_t\r)^2}{2\sigma^2\tau_t}\r\}
      \leq \frac{1}{\sqrt{2\pi}\sigma\sqrt{\tau_t}}\exp\l\{-\frac{\l(\frac{\sigma^2}{2}\r)^2}{2\sigma^2\tau_t}\r\} \nonumber \\
&\leq \frac{1}{\sqrt{2\pi}\sigma\sqrt{\tau_t}}\exp\l\{-\frac{\sigma^2}{8\tau_t}\r\}
      \leq \frac{8\sqrt{\tau_t}}{\sqrt{2\pi}\sigma^3e}.
\end{align}
for any $z\in(0,\frac{z^0}{2}]$, and any $\tau_t\in(0,1]$, where $\phi$ is the probability density function of the standard normal distribution.
Remark that the last inequality in (\ref{eq-thm1-6}) is due to the fact that $ue^{-u}\leq e^{-1}$ holds for any $u\geq0$.
Thus, noting that
\begin{align*}
\calL^z(t,x,\sigma^2)
&= e^{x_z}\Phi(d^+_{\rho z})-Ke^{-r\tau_u}\Phi(d^-_{\rho z})-e^x\Phi(d^+)+Ke^{-r\tau_u}\Phi(d^-)+e^x\Phi(d^+)(1-e^{\rho z}) \\
&= e^{x_z}(\Phi(d^+_{\rho z})-\Phi(d^+))-Ke^{-r\tau_u}(\Phi(d^-_{\rho z})-\Phi(d^-)),
\end{align*}
we obtain
\begin{align}\label{eq-thm1-6-1}
\Bigg|\int_0^{\frac{z^0}{2}}\calL^z(t,x,\sigma^2)\nu(dz)\Bigg|
&\leq \int_0^{\frac{z^0}{2}}e^{x_z}(\Phi(d^+)-\Phi(d^+_{\rho z}))\nu(dz)+K\int_0^{\frac{z^0}{2}}(\Phi(d^-)-\Phi(d^-_{\rho z}))\nu(dz) \nonumber \\
&\leq \int_0^{\frac{z^0}{2}}e^x\phi(d^+_{\rho z})(d^+-d^+_{\rho z})\nu(dz)+K\int_0^{\frac{z^0}{2}}\phi(d^-_{\rho z})(d^--d^-_{\rho z})\nu(dz) \nonumber \\
&\leq \int_0^{\frac{z^0}{2}}\l(e^x\phi(d^+_{\rho z})+K\phi(d^-_{\rho z})\r)\frac{|\rho|z}{\sigma\sqrt{\tau_t}}\nu(dz) \nonumber \\
&\leq (e^x+K)\frac{8\sqrt{\tau_t}|\rho|}{\sqrt{2\pi}\sigma^3e}\int_0^{\frac{z^0}{2}}z\nu(dz)
      \leq (e^x+K)\frac{8\sqrt{\tau_t}|\rho|}{\sqrt{2\pi}\sigma^3e}\int_0^{\infty}z\nu(dz).
\end{align}
From the view of (\ref{nu-cond2}), the first term of (\ref{eq-thm1-2}) is less than $C\sqrt{\tau_t}$ for some $C>0$.

Secondly, we prove that the second term of (\ref{eq-thm1-2}) has the same evaluation as the first term for the case of (\ref{eq-thm1-4}).
Remark that $z^0\vee\olz=z^0$ and
\begin{align}\label{eq-thm1-7}
\Bigg|\int_{\frac{z^0}{2}}^{z^0}\calL^z(t,x,\sigma^2)\nu(dz)\Bigg|
&\leq \int_{\frac{z^0}{2}}^{z^0}e^{x_z}(\Phi(d^+)-\Phi(d^+_{\rho z}))\nu(dz)+K\int_{\frac{z^0}{2}}^{z^0}(\Phi(d^-)-\Phi(d^-_{\rho z}))\nu(dz) \nonumber \\
&= \int_{\frac{z^0}{2}}^{z^0}e^{x_z}\int_{d^+_{\rho z}}^{d^+}\phi(\vt)d\vt\nu(dz)+K\int_{\frac{z^0}{2}}^{z^0}\int_{d^-_{\rho z}}^{d^-}\phi(\vt)d\vt\nu(dz).
\end{align}
Now, we evaluate the integrations in the first and second terms on the right hand side of (\ref{eq-thm1-7}) at once by using Assumption \ref{ass1}.
To this end, we denote
\begin{align*}
h^\pm(\vt) &:= z^0+\frac{\pm\frac{\sigma^2}{2}\tau_t-\sigma\sqrt{\tau_t}\vt}{|\rho|}, \\
d^\pm_0    &:= d^\pm_{\rho z^0}   = \frac{x-\log K+r\tau_t+\rho z^0\pm\frac{\sigma^2}{2}\tau_t}{\sigma\sqrt{\tau_t}}=\pm\frac{\sigma}{2}\sqrt{\tau_t}, \\
d^\pm_2    &:= d^\pm_{\rho z^0/2} = \frac{x-\log K+r\tau_t+\rho\frac{z^0}{2}\pm\frac{\sigma^2}{2}\tau_t}{\sigma\sqrt{\tau_t}}>0,
\end{align*}
where double sign corresponds.
Note that the function $\gamma(z)=z^{-\frac{3}{2}}\vee z^{-\frac{1}{2}}$ defined in (\ref{ass1-1}) is decreasing, and $h^\pm(\vt)\geq z^0/2$ for any $\vt\in[d^\pm_0,d^\pm_2]$.
We have then
\begin{align}\label{eq-thm1-7-2}
\int_{\frac{z^0}{2}}^{z^0}\int_{d^\pm_{\rho z}}^{d^\pm}\phi(\vt)d\vt\nu(dz)
&=    \int_{d^\pm_0}^{d^\pm_2}\int_{h^\pm(\vt)}^{z^0}\nu(dz)\phi(\vt)d\vt+\int_{d^\pm_2}^{d^\pm}\int_{\frac{z^0}{2}}^{z^0}\nu(dz)\phi(\vt)d\vt \nonumber \\
&\leq C^\nu_0\gamma\l(\frac{z^0}{2}\r)\Bigg\{\int_{d^\pm_0}^{d^\pm_2}(z^0-h^\pm(\vt))\phi(\vt)d\vt+\frac{z^0}{2}\int_{d^\pm_2}^{d^\pm}\phi(\vt)d\vt\Bigg\} \nonumber \\
&\leq C^\nu_0\gamma\l(\frac{\olz}{2}\r)\Bigg\{\int_{-\infty}^\infty\frac{\frac{\sigma^2}{2}\tau_t+\sigma\sqrt{\tau_t}|\vt|}{|\rho|}\phi(\vt)d\vt
      +\frac{z^0}{2}\int_{d^\pm_2}^{d^\pm}\phi(\vt)d\vt\Bigg\} \nonumber \\
&\leq C^\nu_0\gamma\l(\frac{\olz}{2}\r)\Bigg\{\frac{1}{|\rho|}\l(\frac{\sigma^2}{2}\tau_t+\sigma\sqrt{\frac{2\tau_t}{\pi}}\r)+\frac{z^0}{2}\int_{d^\pm_2}^{d^\pm}\phi(\vt)d\vt\Bigg\},
\end{align}
where $\olz=\frac{2\sigma^2}{|\rho|}$ and double sign corresponds.
As for the second term in (\ref{eq-thm1-7-2}), a similar argument to (\ref{eq-thm1-6}) implies
\begin{align}\label{eq-thm1-7-3}
\frac{z^0}{2}\int_{d^\pm_2}^{d^\pm}\phi(\vt)d\vt
&\leq \frac{z^0}{2}\phi(d^\pm_2)(d^\pm-d^\pm_2)
      \leq \frac{x-\log K+r\tau_t}{2|\rho|}\phi\l(\frac{x-\log K+r\tau_t}{4\sigma\sqrt{\tau_t}}\r)\frac{x-\log K+r\tau_t}{2\sigma\sqrt{\tau_t}} \nonumber \\
&\leq \frac{1}{\sqrt{2\pi}}\exp\l\{-\frac{(x-\log K+r\tau_t)^2}{32\sigma^2\tau_t}\r\}\frac{(x-\log K+r\tau_t)^2}{4|\rho|\sigma\sqrt{\tau_t}}
      \leq \frac{8\sigma\sqrt{\tau_t}}{\sqrt{2\pi}|\rho|e}.
\end{align}
Remark that the second inequality is derived from the fact that
\[
d^\pm_2 \geq d^-_2 \geq \frac{\frac{x-\log K+r\tau_t}{2}-\frac{\sigma^2}{2}}{\sigma\sqrt{\tau_t}} \geq \frac{x-\log K+r\tau_t}{4\sigma\sqrt{\tau_t}}>0
\]
holds by (\ref{eq-thm1-4}).
From (\ref{eq-thm1-7}) -- (\ref{eq-thm1-7-3}), the second term of (\ref{eq-thm1-2}) is less than $C\sqrt{\tau_t}$ for some $C>0$.

Lastly we discuss the third term of (\ref{eq-thm1-2}).
Since $\Phi(d^+_{\rho z})\geq\Phi(d^-_{\rho z})$, we have
\begin{align}\label{eq-thm1-8-2}
& \int_{z^0}^\infty\l|\calL^z(t,x,\sigma^2)-\l(Ke^{-r\tau_t}\Phi(d^-)-e^{x_z}\Phi(d^+)\r)\r|\nu(dz) \nonumber \\
&=    \int_{z^0}^\infty\l|e^{x_z}\Phi(d^+_{\rho z})-Ke^{-r\tau_u}\Phi(d^-_{\rho z})\r|\nu(dz)
      \leq (e^x+K)\int_{z^0}^\infty\Phi(d^+_{\rho z})\nu(dz).
\end{align}
In addition, a similar calculation to (\ref{eq-thm1-7-2}) provides that
\begin{align}\label{eq-thm1-9}
\int_{z^0}^\infty\Phi(d^+_{\rho z})\nu(dz)
&=    \int_{z^0}^\infty\int_{-\infty}^{d^+_{\rho z}}\phi(\vt)d\vt\nu(dz)
      = \int_{-\infty}^{d^+_0}\int_{z^0}^{h^+(\vt)}\nu(dz)\phi(\vt)d\vt \nonumber \\
&\leq \frac{C^\nu_0}{|\rho|}\gamma(z^0)\int_{-\infty}^{d^+_0}\l(\frac{\sigma^2\tau_t}{2}-\sigma\sqrt{\tau_t}\vt\r)\phi(\vt)d\vt \nonumber \\
&\leq \frac{C^\nu_0}{|\rho|}\gamma(\olz)\l(\frac{\sigma^2\tau_t}{2}-\sigma\sqrt{\tau_t}\int_{-\infty}^0\vt\phi(\vt)d\vt\r)
      \leq \frac{C^\nu_0}{|\rho|}\gamma(\olz)\l(\frac{\sigma^2\tau_t}{2}+\sigma\sqrt{\frac{\tau_t}{2\pi}}\r).
\end{align}
As a result, the third term of (\ref{eq-thm1-2}) has the same evaluation as the first and second terms, which implies that (\ref{eq-thm1-1}) follows when $x-\log K\geq2\sigma^2$.

\noindent{\it Step 2.} \ 
We show (\ref{eq-thm1-1}) for the case where $|x-\log K|<2\sigma^2$.
To this end, we prove that there is a $C>0$ such that
\begin{equation}\label{eq-thm1-10}
\Bigg|\int_0^{z^0\vee\olz}\calL^z(t,x,\sigma^2)\nu(dz)\Bigg|<C,
\end{equation}
and
\begin{equation}\label{eq-thm1-11}
\int_{z^0\vee\olz}^\infty\l|\calL^z(t,x,\sigma^2)-\l(Ke^{-r\tau_t}\Phi(d^-)-e^{x_z}\Phi(d^+)\r)\r|\nu(dz)<C\sqrt{\tau_t}.
\end{equation}
Remark that, although it is enough to see that the left hand side of (\ref{eq-thm1-11}) is less than some constant $C>0$ to prove Theorem \ref{thm1},
we shall give an evaluation with higher order as above.

We show (\ref{eq-thm1-11}) firstly.
From the view of (\ref{eq-thm1-8-2}), we have only to evaluate
\[
\int_{\olz}^\infty\Phi(d^+_{\rho z})\nu(dz),
\]
since $\olz\leq z^0\vee\olz$.
By the same manner as (\ref{eq-thm1-9}), we have
\begin{align}\label{eq-thm1-12}
\int_{\olz}^\infty\Phi(d^+_{\rho z})\nu(dz)
&=    \int_{\olz}^\infty\int_{-\infty}^{d^+_{\rho z}}\phi(\vt)d\vt\nu(dz)
      = \int_{-\infty}^{\old^+}\int_{\olz}^{h^+(\vt)}\nu(dz)\phi(\vt)d\vt \nonumber \\
&\leq C^\nu_0\gamma(\olz)\int_{-\infty}^{\old^+}(h^+(\vt)-\olz)\phi(\vt)d\vt \nonumber \\
&=    C^\nu_0\gamma(\olz)\int_{-\infty}^{\old^+}\l\{z^0-\olz+\frac{1}{|\rho|}\l(\frac{\sigma^2\tau_t}{2}-\sigma\sqrt{\tau_t}\vt\r)\r\}\phi(\vt)d\vt \nonumber \\
&\leq \frac{C^\nu_0}{|\rho|}\gamma(\olz)\int_{-\infty}^\infty\l(r\tau_t+\frac{\sigma^2\tau_t}{2}+\sigma\sqrt{\tau_t}|\vt|\r)\phi(\vt)d\vt \nonumber \\
&\leq \frac{C^\nu_0}{|\rho|}\gamma(\olz)\l(r\tau_t+\frac{\sigma^2\tau_t}{2}+\sigma\sqrt{\frac{2\tau_t}{\pi}}\r),
\end{align}
where
\[
\old^+:=d^+_{\rho\olz}=\frac{x-\log K+r\tau_t+\rho\olz+\frac{\sigma^2}{2}\tau_t}{\sigma\sqrt{\tau_t}}.
\]
Note that the second inequality in (\ref{eq-thm1-12}) is derived from the fact that $z^0-\olz\leq r\tau_t/|\rho|$ holds.
As a consequence, we obtain
\[
\int_{z^0\vee\olz}^\infty\l|\calL^z(t,x,\sigma^2)-\l(Ke^{-r\tau_t}\Phi(d^-)-e^{x_z}\Phi(d^+)\r)\r|\nu(dz)
\leq (e^x+K)\frac{C^\nu_0}{|\rho|}\gamma(\olz)\l(r\tau_t+\frac{\sigma^2\tau_t}{2}+\sigma\sqrt{\frac{2\tau_t}{\pi}}\r),
\]
from which (\ref{eq-thm1-11}) follows.

Next, we aim to see (\ref{eq-thm1-10}), whose left hand side is decomposed as follows:
\begin{align}\label{eq-thm1-13}
&\Bigg|\int_0^{z^0\vee\olz}\calL^z(t,x,\sigma^2)\nu(dz)\Bigg| \nonumber \\
&=    \Bigg|\int_0^{z^0\vee\olz}\l\{e^{x_z}(\Phi(d^+)-\Phi(d^+_{\rho z}))-Ke^{-r\tau_t}(\Phi(d^-)-\Phi(d^-_{\rho z}))\r\}\nu(dz)\Bigg| \nonumber \\
&\leq e^x\int_0^{z^0\vee\olz}(1-e^{\rho z})(\Phi(d^+)-\Phi(d^+_{\rho z}))\nu(dz)+\l|e^x-Ke^{-r\tau_t}\r|\int_0^{z^0\vee\olz}(\Phi(d^+)-\Phi(d^+_{\rho z}))\nu(dz) \nonumber \\
&\hspace{5mm} +Ke^{-r\tau_t}\l|\int_0^{z^0\vee\olz}\l\{(\Phi(d^+)-\Phi(d^+_{\rho z}))-(\Phi(d^-)-\Phi(d^-_{\rho z}))\r\}\nu(dz)\r|.
\end{align}
Note that the first term of (\ref{eq-thm1-13}) is bounded, since
\[
\int_0^{z^0\vee\olz}(1-e^{\rho z})(\Phi(d^+)-\Phi(d^+_{\rho z}))\nu(dz) \leq \int_0^{z^0\vee\olz}(1-e^{\rho z})\nu(dz) \leq \int_0^\infty(1-e^{\rho z})\nu(dz).
\]
In addition, the last term of (\ref{eq-thm1-13}) is evaluated as
\begin{align*}
&\l|\int_0^{z^0\vee\olz}\l\{(\Phi(d^+)-\Phi(d^+_{\rho z}))-(\Phi(d^-)-\Phi(d^-_{\rho z}))\r\}\nu(dz)\r| \\
&=    \l|\int_0^{z^0\vee\olz}\l\{(\Phi(d^+)-\Phi(d^-))-(\Phi(d^+_{\rho z})-\Phi(d^-_{\rho z}))\r\}\nu(dz)\r| \\
&=    \l|\int_0^{z^0\vee\olz}\l\{\int_{d^-}^{d^+}\phi(\vt)d\vt-\int_{d^-_{\rho z}}^{d^+_{\rho z}}\phi(\vt)d\vt\r\}\nu(dz)\r| \\
&=    \l|\int_0^{z^0\vee\olz}\int_{d^-}^{d^+}\l\{\phi(\vt)-\phi\l(\vt+\frac{\rho z}{\sigma\sqrt{\tau_t}}\r)\r\}d\vt\nu(dz)\r| \\
&=    \l|\int_0^{z^0\vee\olz}\int_{d^-}^{d^+}\int_{\vt+\frac{\rho z}{\sigma\sqrt{\tau_t}}}^{\vt}\phi^\prime(\zeta)d\zeta d\vt\nu(dz)\r|
      \leq \int_0^{z^0\vee\olz}\int_{d^-}^{d^+}\frac{|\rho|z}{\sqrt{2\pi e}\sigma\sqrt{\tau_t}}d\vt\nu(dz) \\
&=    \int_0^{z^0\vee\olz}\frac{|\rho|z}{\sqrt{2\pi e}}\nu(dz)
      \leq \frac{|\rho|}{\sqrt{2\pi e}}\int_0^\infty z \nu(dz).
\end{align*}
Note that the first inequality is derived from that $|\phi^\prime(\zeta)|\leq\frac{1}{\sqrt{2\pi e}}$ for any $\zeta\in\bbR$.

The second term of (\ref{eq-thm1-13}) remains to be evaluated.
First of all, we calculate
\[
(e^x-Ke^{-r\tau_t})\int_0^{\frac{z^0}{2}}(\Phi(d^+)-\Phi(d^+_{\rho z}))\nu(dz)
\]
for the case where $x-\log K+r\tau_t>0$.
For any $z\in(0,\frac{z^0}{2}]$, we have
\[
d^+_{\rho z}=\frac{x-\log K+r\tau_t+\rho z+\frac{1}{2}\sigma^2\tau_t}{\sigma\sqrt{\tau_t}} \geq \frac{x-\log K+r\tau_t+\sigma^2\tau_t}{2\sigma\sqrt{\tau_t}} > 0,
\]
which implies that
\[
\Phi(d^+)-\Phi(d^+_{\rho z}) \leq \phi(d^+_{\rho z})(d^+-d^+_{\rho z}) = \phi(d^+_{\rho z})\frac{|\rho|z}{\sigma\sqrt{\tau_t}}.
\]
Now, since $ue^{-u^2}\leq\frac{1}{\sqrt{2e}}$ holds for any $u>0$, we have
\begin{align*}
\frac{\phi(d^+_{\rho z})}{\sigma\sqrt{\tau_t}}
&=    \frac{1}{\sqrt{2\pi}\sigma\sqrt{\tau_t}}\exp\l\{-\frac{(x-\log K+r\tau_t+\rho z+\frac{\sigma^2}{2}\tau_t)^2}{2\sigma^2\tau_t}\r\} \\
&\leq \frac{1}{\sqrt{2\pi}\sigma\sqrt{\tau_t}}\exp\l\{-\frac{\l(\frac{x-\log K+r\tau_t}{2}\r)^2}{2\sigma^2\tau_t}\r\} \\
&\leq \frac{1}{\sqrt{2\pi}\sigma\sqrt{\tau_t}}\frac{2\sqrt{2}\sigma\sqrt{\tau_t}}{\sqrt{2e}(x-\log K+r\tau_t)}
      \leq \sqrt{\frac{2}{\pi e}}\frac{1}{x-\log K+r\tau_t}
\end{align*}
for any $z\in(0,\frac{z^0}{2}]$.
On the other hand, denoting $y := x-\log K+r\tau_t<2\sigma^2+r\tau_t$, we have
\begin{equation}\label{eq-thm1-14}
\frac{e^x-Ke^{-r\tau_t}}{x-\log K+r\tau_t} = Ke^{-r\tau_t}\frac{e^y-1}{y} \leq Ke^{-r\tau_t}\frac{\exp\{2\sigma^2+r\tau_t\}-1}{2\sigma^2+r\tau_t} \leq \frac{Ke^{2\sigma^2}}{2\sigma^2}.
\end{equation}
As a result, we obtain
\begin{align}\label{eq-thm1-14-2}
&(e^x-Ke^{-r\tau_t})\int_0^{\frac{z^0}{2}}(\Phi(d^+)-\Phi(d^+_{\rho z}))\nu(dz) \nonumber \\
&\leq (e^x-Ke^{-r\tau_t})\int_0^{\frac{z^0}{2}}\frac{\phi(d^\pm_{\rho z})}{\sigma\sqrt{\tau_t}}|\rho|z\nu(dz)
      \leq \frac{e^x-Ke^{-r\tau_t}}{x-\log K+r\tau_t}\sqrt{\frac{2}{\pi e}}|\rho|\int_0^\infty z\nu(dz) \nonumber \\
&\leq \frac{Ke^{2\sigma^2}}{2\sigma^2}\sqrt{\frac{2}{\pi e}}|\rho|\int_0^\infty z\nu(dz).
\end{align}

Next, we evaluate
\begin{equation}\label{eq-thm1-14-3}
(e^x-Ke^{-r\tau_t})\int_{\frac{z^0}{2}}^{z^0\vee\olz}(\Phi(d^+)-\Phi(d^+_{\rho z}))\nu(dz)
\end{equation}
for the case where $x-\log K+r\tau_t>0$.
Since $0<\Phi(d^+)-\Phi(d^+_{\rho z})<1$, it suffices to evaluate $\nu([\frac{z^0}{2},z^0\vee\olz])$.
When $\frac{z^0}{2}<1$, Assumption \ref{ass1} ensures that
\begin{align*}
\nu\l(\l[\frac{z^0}{2},z^0\vee\olz\r]\r)
&\leq C^\nu_0\l\{\int_{\frac{z^0}{2}}^{1\wedge(z^0\vee\olz)}z^{-\frac{3}{2}}dz+\int_{1\wedge(z^0\vee\olz)}^{z^0\vee\olz}z^{-\frac{1}{2}}dz\r\} \\
&=    C^\nu_0\l\{\frac{2\sqrt{2|\rho|}}{\sqrt{x-\log K+r\tau_t}}-\frac{2}{\sqrt{1\wedge(z^0\vee\olz)}}+2\sqrt{z^0\vee\olz}-2\sqrt{1\wedge(z^0\vee\olz)}\r\} \\
&\leq C^\nu_0\l\{\frac{2\sqrt{2|\rho|}}{\sqrt{x-\log K+r\tau_t}}+2\sqrt{z^0\vee\olz}\r\}.
\end{align*}
We have then
\begin{align}\label{eq-thm1-15}
&(e^x-Ke^{-r\tau_t})\int_{\frac{z^0}{2}}^{z^0\vee\olz}(\Phi(d^+)-\Phi(d^+_{\rho z}))\nu(dz) \nonumber \\
&\leq C^\nu_0(e^x-Ke^{-r\tau_t})\l\{\frac{2\sqrt{2|\rho|}}{\sqrt{x-\log K+r\tau_t}}+2\sqrt{z^0\vee\olz}\r\} \nonumber \\
&\leq C^\nu_0Ke^{2\sigma^2}\l\{\frac{2\sqrt{2|\rho|}}{\sqrt{2\sigma^2+r\tau_t}}+2\sqrt{z^0\vee\olz}\r\}
      \leq 2C^\nu_0Ke^{2\sigma^2}\l\{\frac{\sqrt{|\rho|}}{\sigma}+\sqrt{\frac{2\sigma^2+r}{|\rho|}}\r\},
\end{align}
since $z^0\vee\olz\leq(2\sigma^2+r)/|\rho|$ holds, and similar calculations to (\ref{eq-thm1-14}) provide
\begin{equation}\label{eq-thm1-15-2}
\frac{e^x-Ke^{-r\tau_t}}{\sqrt{x-\log K+r\tau_t}} \leq \frac{Ke^{2\sigma^2}}{\sqrt{2\sigma^2+r\tau_t}} \ \ \ \mbox{ and } \ \ \ e^x-Ke^{-r\tau_t} \leq Ke^{2\sigma^2}.
\end{equation}

Moreover, when $\frac{z^0}{2}\geq1$, we have
\[
\nu\l(\l[\frac{z^0}{2},z^0\vee\olz\r]\r) \leq C^\nu_0\int_{\frac{z^0}{2}}^{z^0\vee\olz}z^{-\frac{1}{2}}dz \leq 2C^\nu_0\sqrt{z^0\vee\olz}.
\]
By using (\ref{eq-thm1-15-2}), we have
\begin{equation}\label{eq-thm1-16}
(e^x-Ke^{-r\tau_t})\int_{\frac{z^0}{2}}^{z^0\vee\olz}(\Phi(d^+)-\Phi(d^+_{\rho z}))\nu(dz) \leq 2C^\nu_0Ke^{2\sigma^2}\sqrt{\frac{2\sigma^2+r}{|\rho|}}.
\end{equation}
From (\ref{eq-thm1-15}) and (\ref{eq-thm1-16}), we can find a constant $C>0$ such that (\ref{eq-thm1-14-3}) is less than $C$.
Together with the result of (\ref{eq-thm1-14-2}), the second term of (\ref{eq-thm1-13}) is bounded when $x-\log K +r\tau_t>0$.

As the last step to evaluate the second term of (\ref{eq-thm1-13}), we treat the case where $x-\log K +r\tau_t\leq0$ and $x-\log K>-2\sigma^2$.
In this case, $z^0$ takes a nonpositive value, so that $z^0\vee\olz=\olz$ holds.
The following shows that $\phi(d^+_{\rho z})$ is bounded as a function on $z\in(0,\olz]$:
For any $z\in(0,\olz]$, we have
\begin{align*}
\phi(d^+_{\rho z})
&=    \frac{1}{\sqrt{2\pi}}\exp\l\{-\frac{(x-\log K+r\tau_t+\rho z+\frac{\sigma^2}{2}\tau_t)^2}{2\sigma^2\tau_t}\r\} \\
&=    \frac{1}{\sqrt{2\pi}}\exp\l\{-\frac{(x-\log K+r\tau_t+\rho z)^2}{2\sigma^2\tau_t}-\frac{x-\log K+r\tau_t+\rho z}{2}-\frac{\sigma^2\tau_t}{8}\r\} \\
&\leq \frac{1}{\sqrt{2\pi}}\exp\l\{-\frac{(x-\log K+r\tau_t)^2}{2\sigma^2\tau_t}\r\}\exp\l\{\frac{|x-\log K|+|\rho|\olz}{2}\r\} \\
&\leq \frac{1}{\sqrt{2\pi}}\exp\l\{-\frac{(x-\log K+r\tau_t)^2}{2\sigma^2\tau_t}\r\}e^{2\sigma^2} \\
&\leq \frac{1}{\sqrt{2\pi}}\frac{1}{\sqrt{2e}}\frac{\sqrt{2}e^{2\sigma^2}\sigma\sqrt{\tau_t}}{|x-\log K+r\tau_t|}
      = \frac{e^{2\sigma^2}\sigma\sqrt{\tau_t}}{\sqrt{2\pi e}|x-\log K+r\tau_t|},
\end{align*}
since $|x-\log K|<2\sigma^2$ and $ue^{-u^2}\leq\frac{1}{\sqrt{2e}}$ for any $u>0$.
Thus, we obtain
\begin{align}\label{eq-thm1-17}
&(Ke^{-r\tau_t}-e^x)\int_0^{\olz}(\Phi(d^+)-\Phi(d^+_{\rho z}))\nu(dz) \nonumber \\
&=    (Ke^{-r\tau_t}-e^x)\int_0^{\olz}\int^{d^+}_{d^+_{\rho z}}\phi(\vt)d\vt\nu(dz) \nonumber \\
&\leq (Ke^{-r\tau_t}-e^x)\frac{e^{2\sigma^2}\sigma\sqrt{\tau_t}}{\sqrt{2\pi e}|x-\log K+r\tau_t|}\int_0^{\olz}\frac{|\rho|z}{\sigma\sqrt{\tau_t}}\nu(dz) \nonumber \\
&\leq \frac{e^{2\sigma^2}}{\sqrt{2\pi e}}\frac{e^{2\sigma^2+x}}{2\sigma^2}|\rho|\int_0^\infty z\nu(dz)
      = \frac{e^{4\sigma^2+x}|\rho|}{2\sqrt{2\pi e}\sigma^2}\int_0^\infty z\nu(dz).
\end{align}
Remark that the same sort of argument as (\ref{eq-thm1-14}) implies
\[
\frac{Ke^{-r\tau_t}-e^x}{\log K-r\tau_t-x} \leq \frac{e^{2\sigma^2+x}}{2\sigma^2}
\]
when $x-\log K+r\tau_t\leq0$ and $x-\log K>-2\sigma^2$.
Hence, (\ref{eq-thm1-17}) provides the boundedness of the second term of (\ref{eq-thm1-13}) as a function on $\tau_t$ in this case.

Consequently, (\ref{eq-thm1-10}) holds when $|x-\log K|<2\sigma^2$, from which (\ref{eq-thm1-1}) follows for the case where $|x-\log K|<2\sigma^2$.
Together with Step 1, the proof of Theorem \ref{thm1} is completed.
\fin

\subsection{Proof of Theorem \ref{thm2}}
From the views of Proposition \ref{prop1} and Step 1 in the proof of Theorem \ref{thm1}, it is enough to see that
\begin{equation}\label{eq-thm2-1}
\Bigg|\int_{z^0}^\infty\l\{\calL^z(t,x,\sigma^2)-\l(Ke^{-r\tau_t}-e^{x_z}\r)\r\}\nu(dz)\Bigg|\leq C\tau_t^{\frac{1}{2}}
\end{equation}
for some $C>0$ when $x-\log K\geq2\sigma^2$.
Note that
\[
1-\Phi(\vt)\leq \frac{1}{\sqrt{2\pi}\vt}
\]
holds true for any $\vt>0$, and
\[
d^\pm \geq \frac{2\sigma^2\pm\frac{\sigma^2}{2}\tau_t}{\sigma\sqrt{\tau_t}} \geq \frac{3\sigma}{2\sqrt{\tau_t}},
\]
since $x-\log K\geq2\sigma^2$.
Thus, we have
\[
1-\Phi(d^\pm) \leq \frac{1}{\sqrt{2\pi}d^\pm} \leq \sqrt{\frac{2}{\pi}}\frac{\sqrt{\tau_t}}{3\sigma},
\]
which provides that
\begin{align}\label{eq-thm2-2}
& \int_{z^0}^\infty|\calL^z(t,x,\sigma^2)-(Ke^{-r\tau_t}-e^{x_z})|\nu(dz) \nonumber \\
&=    \int_{z^0}^\infty\l|e^{x_z}\l(\Phi(d^+_{\rho z})-\Phi(d^+)+1\r)-Ke^{-r\tau_u}\l(\Phi(d^-_{\rho z})-\Phi(d^-)+1\r)\r|\nu(dz) \nonumber \\
&\leq \int_{z^0}^\infty\l\{e^x\Phi(d^+_{\rho z})+(e^x+K)\sqrt{\frac{2}{\pi}}\frac{\sqrt{\tau_t}}{3\sigma}+K\Phi(d^-_{\rho z})\r\}\nu(dz) \nonumber \\
&\leq (e^x+K)\int_{z^0}^\infty\Phi(d^+_{\rho z})\nu(dz)+(e^x+K)\sqrt{\frac{2}{\pi}}\frac{\sqrt{\tau_t}}{3\sigma}\nu([\olz,\infty)) \nonumber \\
&\leq (e^x+K)\l\{\frac{C^\nu_0}{|\rho|}\gamma(\olz)\l(\frac{\sigma^2\tau_t}{2}+\sigma\sqrt{\frac{\tau_t}{2\pi}}\r)
      +\sqrt{\frac{2}{\pi}}\frac{\sqrt{\tau_t}}{3\sigma}\nu([\olz,\infty))\r\}.
\end{align}
Note that the second inequality is due to $\Phi(d^+_{\rho z})\geq\Phi(d^-_{\rho z})$ and $z^0>\olz$, and the last inequality is derived from (\ref{eq-thm1-9}).
In addition, $\nu([\olz,\infty))$ is evaluated by using (\ref{ass1-1}) as follows:
\[
\nu([\olz,\infty))
\leq C^\nu_0\l\{\int_{\olz}^{1\vee\olz}z^{-\frac{3}{2}}dz+\int_{1\vee\olz}^\infty e^{-C^\nu_1z}dz\r\}
\leq C^\nu_0\l\{\frac{2}{\sqrt{\olz}}+\frac{1}{C^\nu_1}\r\}
=    C^\nu_0\l\{\frac{\sqrt{2|\rho|}}{\sigma}+\frac{1}{C^\nu_1}\r\}.
\]
Together with (\ref{eq-thm2-2}), we obtain (\ref{eq-thm2-1}).
This completes the proof Theorem \ref{thm2}.
\fin

%
%
\setcounter{equation}{0}
\section{Conclusions}
Approximate expressions of call option prices for the BNS model have been developed in this paper.
As indicated by numerical results in Subsection 3.2, our approximate expressions are sufficiently effective for ITM and around ATM options with a short maturity.
Developments of approximation methods for implied volatilities and calibration procedures for model parameters are significant problems
as applications of our approximations, but these are left to future works.

\begin{center}
{\bf Acknowledgments}
\end{center}
Takuji Arai gratefully acknowledges the financial support of the MEXT Grant in Aid for Scientific Research (C) No.18K03422.



\begin{thebibliography}{9999}
\bibitem{Alos12}
E. Al\`os, A decomposition formula for option prices in the Heston model and applications to option pricing approximation, Finance \& Stochastics, 16 (2012), pp.403-422.
\bibitem{ADV15}
E. Al\`os, R. De Santiago and J. Vives, Calibration of stochastic volatility models via second-order approximation: the Heston case,
International Journal of Theoretical and Applied Finance, 18 (2015), 1550036.
\bibitem{A}
T. Arai,  Al\`os type decomposition formula for Barndorff-Nielsen and Shephard model, to appear in Journal of Stochastic Analysis (2021).
\bibitem{AIS-BNS}
T. Arai,Y, Imai and R. Suzuki, Local risk-minimization for Barndorff-Nielsen and Shephard models, Finance \& Stochastics, 21 (2017), pp.551-592.
\bibitem{AS}
T. Arai and R, Suzuki, Local risk-minimization for L\'evy markets, International Journal of Financial Engineering, 2 (2015), 1550015.
\bibitem{BNS1}
O.E. Barndorff-Nielsen and N. Shephard, Modelling by L\'evy processes for financial econometrics. In: Barndorff-Nielsen, O.E., Mikosch,T., Resnick, S. (eds.):
L\'evy processes---Theory and Applications, Birkh\"auser, Basel, (2001), pp.283-318.
\bibitem{BNS2}
O.E. Barndorff-Nielsen and N. Shephard, Non-Gaussian Ornstein-Uhlenbeck based models and some of their uses in financial econometrics,
J.R. Statistic. Soc., 63 (2001), pp.167--241.
\bibitem{CT}
R. Cont and P. Tankov, Financial Modeling with Jump Process, Chapman \& Hall, London, 2004.
\bibitem{MPSV18}
R. Merino, J. Posp\`i\v{s}il, T. Sobotka and J. Vives, Decomposition formula for jump diffusion models, International Journal of Theoretical and Applied Finance, 21 (2018), 1850052.
\bibitem{MV17}
R. Merino and J. Vives, Option price decomposition in spot-dependent volatility models and some applications, International Journal of Stochastic Analysis, (2017), 8019498.
\bibitem{NV}
E. Nicolato and E. Venardos, E. Option pricing in stochastic volatility models of the Ornstein-\"Uhlenbeck type, Mathematical Finance, 13 (2003), pp.445-466.
\bibitem{Scho}
W. Schoutens, L\'evy processes in finance: pricing financial derivatives, Wiley, 2003.
\end{thebibliography}
\end{document}